\documentclass[10pt,journal,compsoc]{IEEEtran}

\ifCLASSOPTIONcompsoc
  \usepackage[nocompress]{cite}
\else
  \usepackage{cite}
\fi
\usepackage{amsmath,amssymb,amsfonts}
\usepackage{algorithmic}
\usepackage[linesnumbered,ruled]{algorithm2e}
\usepackage{amsthm}
\usepackage{graphicx}
\usepackage{textcomp}
\usepackage{paralist}
\usepackage{subcaption}
\usepackage{hyperref}
\usepackage{multirow}
\usepackage{url}
\usepackage{setspace}
\usepackage{tabu}
\usepackage{romannum}
\usepackage{flushend}
\usepackage{enumitem}
\usepackage{arydshln}
\usepackage[normalem]{ulem}
\usepackage{threeparttable}
\usepackage{url}
\usepackage{xcolor}
\usepackage{soul}
\newcommand{\yun}[1]{\textcolor{black}{#1}}

\makeatletter
\newcommand{\thickhline}{%
	\noalign {\ifnum 0=`}\fi \hrule height 1pt
	\futurelet \reserved@a \@xhline
}
\newcolumntype{"}{@{\hskip\tabcolsep\vrule width 1pt\hskip\tabcolsep}}
\makeatother

\ifCLASSINFOpdf
\else
\fi
\begin{document}

\title{Emerging App Issue Identification via Online Joint Sentiment-Topic Tracing}
% User Review-Based Emergent App Issue Detection via Improved Online Topic Modeling

\author{Cuiyun~Gao, Jichuan~Zeng, Zhiyuan~Wen, David~Lo, Xin~Xia, Irwin~King,~\IEEEmembership{Fellow,~IEEE,} and~Michael~R.~Lyu,~\IEEEmembership{Fellow,~IEEE}
        % John~Doe,~\IEEEmembership{Fellow,~OSA,}
        % and~Jane~Doe,~\IEEEmembership{Life~Fellow,~IEEE}% <-this % stops a space
\IEEEcompsocitemizethanks{\IEEEcompsocthanksitem C. Gao, J. Zeng, I. King, and M. R. Lyu are with the Department
of Computer Science and Engineering, The Chinese University of Hong Kong, Shatin,
Hong Kong, China.\protect\\
E-mail: \{cygao,jczeng,king,lyu\}@cse.cuhk.edu.hk
\IEEEcompsocthanksitem Z. Wen is with the Department of Computing, The Hong Kong Polytechnic university, Kowloon, Hong Kong, China.\protect\\
E-mail: cszwen@comp.polyu.edu.hk
\IEEEcompsocthanksitem D. Lo is with the School of Information Systems, Singapore Management University, Singapore.\protect\\
Email: davidlo@smu.edu.sg
\IEEEcompsocthanksitem X. Xia is with the Faculty of Information Technology, Monash University, Australia.\protect\\
Email: xin.xia@monash.edu
\IEEEcompsocthanksitem J. Zeng is the corresponding author.
}% <-this % stops a space
\thanks{Manuscript received August 19, 2020.}}

\markboth{Journal of \LaTeX\ Class Files,~Vol.~14, No.~8, August~2015}%
{Shell \MakeLowercase{\textit{et al.}}: Bare Advanced Demo of IEEEtran.cls for IEEE Computer Society Journals}

\IEEEtitleabstractindextext{%
\begin{abstract}
Millions of mobile apps are available in app stores, such as Apple's App Store and Google Play. For a mobile app, it would be increasingly challenging to stand out from the enormous competitors and become prevalent among users. Good user experience and well-designed functionalities are the keys to a successful app. To achieve this, popular apps usually schedule their updates frequently. If we can capture the critical app issues faced by users in a timely and accurate manner, developers can make timely updates, and good user experience can be ensured. There exist prior studies on analyzing reviews 
% real-time 
for detecting emerging app issues. These studies are usually based on topic modeling or clustering techniques. However, the short-length characteristics and sentiment of user reviews have not been considered. In this paper, we propose a novel emerging issue detection approach named MERIT to take into consideration the two aforementioned characteristics. Specifically, we propose an Adaptive Online Biterm Sentiment-Topic (AOBST) model for jointly modeling topics and corresponding sentiments that takes into consideration app versions. Based on the AOBST model, we infer the topics negatively reflected in user reviews for one app version, and automatically interpret the meaning of the topics with most relevant phrases and sentences. Experiments on popular apps from Google Play and Apple's App Store demonstrate the effectiveness of MERIT in identifying emerging app issues, improving the state-of-the-art method by 22.3\% in terms of F1-score. In terms of efficiency, MERIT can return results within acceptable time.
% Moreover, we verify the performance of MERIT on industrial scenario, and can identify \yun{No.} of the 18 issues in one month. 
\end{abstract}

% Note that keywords are not normally used for peer review papers.
}

\maketitle

% is not selected <OR> if conference mode is selected - because compsoc
% conference papers position the abstract like regular (non-compsoc)
\IEEEdisplaynontitleabstractindextext

% \fi
\IEEEpeerreviewmaketitle

\ifCLASSOPTIONcompsoc
\IEEEraisesectionheading{\section{Introduction}\label{sec:introduction}}
\else
\section{Introduction}
\label{sec:introduction}
\fi
% describe the popularity of apps, the importance of emerging issue detection, the characteristics of app reviews. emerging issue detection in app review field has not been comprehensively studied, online topic modeling approach is approved to be useful but it's limited by several factors.
% the contributions of our paper: more dataset and comparison on industry apps, sentiment, topic labeling, topic number auto define, distributed
\IEEEPARstart{M}{obile} apps keep gaining popularity over the last few years. According to Statista~\cite{appnumber}, the global mobile internet user penetration in 2016 has exceeded half the world's population. During the third quarter of 2018, Android users were able to choose from 2.1 million apps, while Apple's App Store\footnote{Apple's App Store is indicated as App Store for simplicity throughout the paper.} provided almost 2 million apps. While users have a large number of products to choose from, the apps are facing immensely fierce competition to survive.

The popular mobile app stores, such as Google Play and App Store, use the star-rating mechanism to gather users' ratings and feedback. The feedback and ratings can impact an app's ranking on these stores, and further influence its discovery and trial. A survey in 2015~\cite{guidetoappreviews} reported that only 15\%$\sim$50\% of the users would consider downloading a low-rated app, while for the high-rated apps, the ratio reached 96\%. Thus, ensuring good user experience and keeping users engaged can help maintain high download numbers and increase benefits to app developers. 

Recent studies~\cite{DBLP:journals/ese/McIlroyAH16, DBLP:conf/cec/LimB13,appranking} showed that frequently-updated apps could benefit in terms of increase in ranking. This is the case since the popular app stores factor in the freshness of an app in the ranking process. Additionally, app updates can also improve user experience. Specifically, McIlroy et al.~\cite{DBLP:journals/ese/McIlroyAH16} found that the rationale behind updates is often related to bug-fixing (63\% of the time), new features (35\%), and feature improvement (30\%). However, not every update can definitely lead to positive user experience and high ranking~\cite{DBLP:conf/sigsoft/MartinSH16}. For example, the updated Android and iOS versions of Skype released in June 2017 received a flood of complaints as the new design removed the key functionality and features available in the older version, such as the visibility of online friends~\cite{skyperedesign}. As a result, its user rating on the App Store plunged from 4.5 to 1.5 stars shortly after the update~\cite{skyperating}. Such situations are not unusual, c.f.~\cite{messengerbad,pokemonbad}, and can cause customer churn and losses to app developers. The losses could be limited if the issues were recognized timely. In this work, we aim at accurately detecting {\em emerging} app issues by analyzing user feedback.

IDEA~\cite{gao2018online} is one of the most recent works that can be directly applied to detect emerging issues/topics\footnote{The topics and issues are semantically equal in this paper.} from user feedback. IDEA takes user reviews distributed in consecutive app versions as input, and outputs emerging app issues in the level of phrases and sentences. A modified online topic modeling approach is utilized to infer topics of the text corpus in consecutive time periods. Finally, IDEA employs a {\em topic labeling} approach to automatically prioritize the phrases/sentences that are semantically representative of the topics. The prioritized phrases/sentences are regarded as descriptions of emerging issues. Although the approach achieves reasonable performance, it has several limitations in accurately detecting emerging app issues as it does not consider the following characteristics of user feedback \yun{and exists inefficiency during topic labeling}:
% (Online Latent Dirichlet Allocation, OLDA in short) 

1) \textit{\textbf{Short-Length Nature of User Feedback:}} User feedbacks are usually short in length, \yun{providing limited context. According to Genc-Nayebi and Abran~\cite{DBLP:journals/jss/Genc-NayebiA17}, the average length of app reviews is 71 characters.}
% , contain\yun{ing} numerous ``noisy'' words (e.g., misspelled words and abbreviations)\yun{.}
% , and are mostly non-informative. According to Chen et al.~\cite{chen2014ar}, only 30\% of the reviews provide informative user opinions for app updates. 
Besides, since the proposed online topic modeling approach in IDEA is built upon LDA (Latent Dirichlet Allocation)~\cite{DBLP:conf/nips/BleiNJ01} and LDA is not considered to work well on short texts~\cite{DBLP:conf/www/YanGLC13,DBLP:conf/sigir/LiWZSM16}, IDEA may fail to accurately capture the topics of user reviews.

2) \textit{\textbf{Sentiment of Topics:}} Emerging issues are generally the issues that negatively impact user experience, such as bugs, or new features requested by users. Reviews corresponding to these issues are usually accompanied by poor ratings. However, current topic-modeling-based approaches do not explicitly distinguish topics based on their sentiment, which may identify the positive ones as emerging and \yun{generate false positives}.
% false positives.

3) \textit{\textbf{Ineffectiveness of the Topic Labeling Approaches:}} The previous topic labeling approaches represent topics~\cite{gao2018online,DBLP:conf/issre/GaoWHZZL15} with representative candidate phrases/sentences based on their similarities with the current topics in terms of topic distributions. Since topic distributions may not well represent the semantic meanings of words~\cite{DBLP:conf/acl/DasZD15,DBLP:conf/coling/JiangSLW16,DBLP:conf/acl/HuT16,DBLP:conf/acl/LiCZM16}, these topic labeling approaches may choose improper phrases/sentences for interpreting the topics. As the topic interpretations directly represent app issues, false emerging issues would be alerted.

% IDEA infers topics on the whole review corpus without user sentiment explicitly incorporated, and therefore can not distinguish negative and positive topics.

% 3) \textit{\textbf{Parameter Tuning:}} Without available ground truth, unsupervised approaches such as topic modeling are preferred for automatic information extraction. However, the \textit{topic number} in topic modeling can easily impact the results of detected emerging issues~\cite{gao2018online,DBLP:conf/icse/PanichellaDOPPL13}, and manually determining the topic number in different datasets and versions is difficult.

% DIVER~\cite{gao2019diver} is another most recent framework for user-review based emerging issue detection and mainly specific to industry scenario. DIVER is not built on topic modeling approach, but utilizes information retrieval approach to cluster the words that increase abnormally in terms of the percentage. Similar to IDEA~\cite{gao2018online}, DIVER does not explicitly involve user sentiment. Besides, since not all the semantically-related words present an increasing trend for the \textit{emerging} topic, only clustering the \textit{abnormal} words may induce false positives.

In this paper, we propose an i\textbf{M}proved \textbf{E}me\textbf{R}ging \textbf{I}ssue de\textbf{T}ection approach, named \textbf{MERIT}, to mitigate these limitations and more accurately detect emerging app issues. Different from the topic modeling approach in IDEA~\cite{gao2018online}, where a topic is a probability distribution over single words, MERIT considers topics over a mixture of \textit{biterms}. Here, a \textit{biterm} is an unordered word-pair co-occurring in a short context. The biterm-based model has been \yun{shown} to be effective in alleviating the data sparseness problem of short texts and significantly enhance the topic learning~\cite{DBLP:conf/www/YanGLC13}. To tackle the second limitation, MERIT distinctly considers sentiment-related prior during topic modeling, and thereby can well distinguish positive and negative topics. The negative topics are adopted for emerging app issue detection. For the third problem, MERIT employs word embedding~\cite{mikolov2013distributed} which has been \yun{shown} to be effective in converting words into their distributed representations~\cite{pennington2014glove,DBLP:conf/icse/GuZ018}, during the topic labeling process.

% MERIT adopts a simple heuristic method to infer the nearly-optimal topic number. Most importantly, 

To evaluate the effectiveness of MERIT, we perform experiments on the same six real-world apps as IDEA~\cite{gao2018online}. Our results show that MERIT can more accurately identify emerging app issues than the baselines, with improvements in precision, recall, and F1-score of 21.0\%, 20.9\%, and 22.3\% respectively. We discover that MERIT can capture more coherent topics (\textit{i.e.}, the top words belonging to one topic are more semantically consistent) from user reviews, focus on the negative topics, and better prioritize phrases/sentences for interpreting topics. We also demonstrate that MERIT can output results with reasonable time cost despite its more complex design than IDEA.

% We also conduct industry experiment based on WeChat, a popular messenger app with over 1 billion monthly active users, using the app issues confirmed by developers as groudtruth. Moreover, with the support of MERIT, developers can track the sentiment changes of topics.

% named AOBTM (Adaptive Online Biterm Topic Modeling)

The main contributions of our work are as follows:
\begin{itemize}
\item We propose a novel online topic modeling approach for detecting emerging app issues. The proposed approach can not only generate more coherent topics but also well distinguish positive and negative topics during analysis of user reviews.

\item We design a novel topic labeling approach based on word embedding techniques to well prioritize phrases/sentences for interpreting the meaning of each topic.

\item We develop MERIT\footnote{\yun{Available at \url{https://github.com/armor-ai/MERIT}.}}, a new tool that can detect emerging app issues from online reviews.
% We develop an interactive tool to efficiently track the sentiments of topics and view the topic summaries along with app versions.

\item We evaluate the effectiveness and efficiency of MERIT on real-world mobile apps.
\end{itemize}

\textbf{Paper structure.} Section~\ref{sec:background} describes the background knowledge and motivation of our work. Section~\ref{sec:overview} presents the methodology we propose for accurate emerging app issue detection. Section~\ref{sec:setup} introduces the experimental setup. Section~\ref{sec:experiment} describes the evaluation results, followed by Section~\ref{sec:limitation} that discusses the limitation of our approach. Section~\ref{sec:literature} presents related studies.
We conclude and mention future work in Section~\ref{sec:conclusion}.

% \begin{inparaenum}[\bfseries RQ1)]
%     \item What is the performance of proposed approach? \par
%     \item What is the contribution of sentiment or word embeddings in the approach?\par 
% % 	\item What are the major ad issues concerned by users? We aim at extracting and ranking ad-related issues in answering this question.\par
%     \item Can we view the sentiment changes of app issues along with app versions?\par
%     \item What is the performance of our approach in the industry scenario?\par
% \end{inparaenum}
\section{Preliminaries}\label{sec:background}
In this section, we present the background knowledge for facilitating readers' understanding, including emerging issue detection, topic modeling, and word embeddings.

\subsection{\yun{Emerging} Issue Detection}
In mainstream topic detection studies~\cite{DBLP:conf/sigir/ChenALC13,DBLP:conf/acl/DiaoJZL12,DBLP:conf/aaai/YanGLXC15,DBLP:journals/www/HuangPWCGZ17}, an event/issue is considered \yun{emerging} if it is (heavily) discussed in current time slice but not previously. The application scenario of these studies is generally targeted at social media platforms, \textit{e.g.}, Twitter and Sina Weibo. However, user discussion on social media and app stores has significant differences. One difference is that the app reviews usually associate with specific app versions, while typical social media contents do not concern with the version concept. Another big difference is that \yun{emerging} event detection for social media is simply dependent on volume of user posted content, regardless of the sentiment associated with it; while for app reviews, user sentiment is one indicator to \yun{emerging} issues~\cite{guzman2014users,chen2014ar}. Thus, simply applying standard \yun{emerging} issue detection methods from the social media field is not optimal for our scenario. In this paper, we define an \yun{emerging} app issue as follows.

\textbf{Definition (\yun{Emerging} App Issue)} An issue reported in user review(s) at a particular time slice is defined as an emerging issue if \yun{its distributions presented marginal fluctuations in previous time  slices}
% never / seldom  discussed 
in previous time slices, corresponds to a \textit{significant increase} in terms of \textit{the \yun{percentage}}
% number 
\textit{of reviews} reporting it, and is \textit{negatively reported by users} in the current time slice.

\begin{figure}[t]
% \vspace{1mm}
    \centering
    \resizebox{\linewidth}{!}{
    \includegraphics{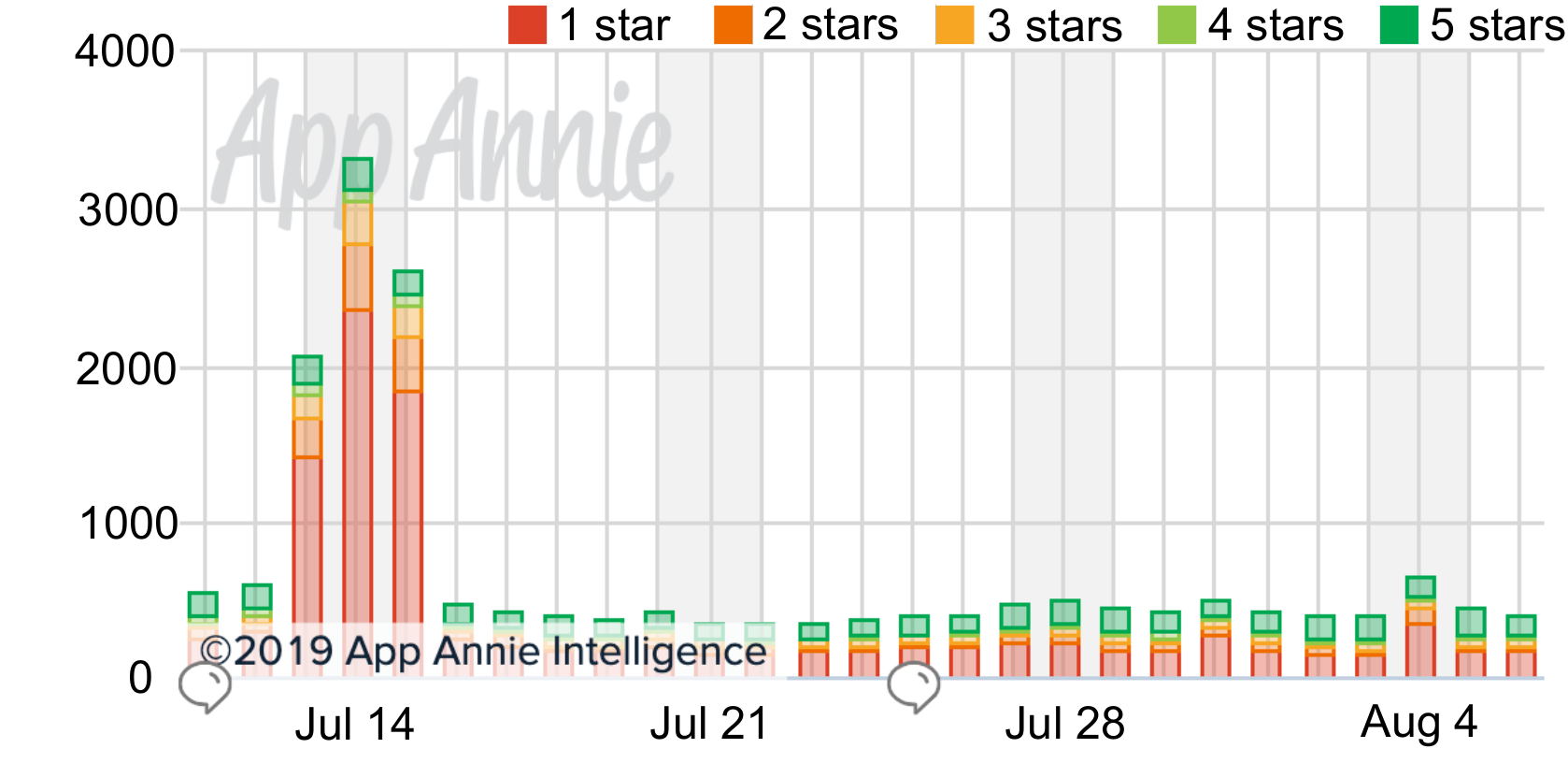}
    }
    \caption{Number of review changes along with time for the Facebook app on Google Play. Different color bars represent the rating distributions with reference shown on the top right. (Statistics from App Annie~\cite{appannie})
    }
    \label{fig:facebook_example}
\end{figure}

For example, Fig.~\ref{fig:facebook_example} illustrates the changes of the number of reviews and user rating distributions over time for Facebook. As can be seen, the number of reviews received on July 14, 2019 is significantly larger than the number of reviews received on July 12, 2019; this is true especially for the one-star reviews (represented by the red bar), which means that an \yun{emerging} issue may exist for the recent update. By checking the detailed user reviews, we find that it was related to a huge redesign of the app in July~\cite{facebookredesign}. With continuous monitoring and accurate identification of \yun{emerging} issues, such problem can be detected in a timely manner. Developers can then be alerted of the need to perform further maintenance activities to ensure good user experience. We also discover that both number of reviews reporting the issue and user rating can indicate the emergence of an issue. Involving review ratings in the analysis can help our tool avoid false positives, i.e., topics that are mentioned in many reviews but do not correspond to important problems that need to be rectified urgently~\cite{DBLP:journals/www/HuangPWCGZ17}.

\subsection{Topic Modeling}
% Topic models~\cite{DBLP:conf/nips/BleiNJ01} are one of the most popular methods for learning unsupervised representations of text~\cite{DBLP:conf/iclr/SrivastavaS17}. 
Topic models~\cite{DBLP:conf/sigir/Hofmann99,DBLP:conf/nips/BleiNJ01} have been proven useful for discovering latent structures in a collection of documents~\cite{DBLP:conf/uai/AhmedX10,DBLP:conf/iclr/SrivastavaS17}. The models capture the co-occurrence of words in the collection under a probabilistic framework by assuming that each topic can be represented by a set of word clusters. In this way, the models can uncover latent semantic structures (\textit{i.e.}, topics) in the documents. Due to the unsupervised nature, the models such as Probabilistic Latent Semantic Analysis (PLSA)~\cite{DBLP:conf/sigir/Hofmann99} and Latent Dirichlet Allocation (LDA)~\cite{DBLP:conf/nips/BleiNJ01} have been widely applied in mining software repositories~\cite{chen2014ar,DBLP:conf/re/GuzmanM14,gao2018online} where labeled datasets are not available in practice. The outputs of the topic models are two matrices: (1) Document-topic matrix, denoted as $\Theta\in \mathbb{R}^{D\times K}$, where $D$ is the number of documents and $K$ is the number of topics. The $i$-th row of the matrix, \textit{i.e.}, $\theta_i\in\mathbb{R}^K$, is a topic distribution vector for the $i$-th document; (2) Topic-word matrix, denoted as $\Phi\in \mathbb{R}^{K\times V}$, where $V$ is the total number of unique words (\textit{i.e.}, vocabulary). The $i$-th row of the matrix, that is, $\phi_i\in\mathbb{R}^V$, is a word distribution vector for the $i$-th topic. Table~\ref{tab:tm_output} shows an example of the output of topic models, with top five words and corresponding probabilities presented for each topic.

\textbf{Latent Dirichlet Allocation (LDA)}~\cite{DBLP:conf/nips/BleiNJ01} 
% is a springboard topic model.
% , which is able to identify potential topic distributions of large corpora. LDA 
assumes that each document consists of a mixture of topics, and each word in documents belongs to one topic. 
% LDA has been proven successful in modeling formal and well-edited documents, such as news reports~\cite{DBLP:conf/nips/BleiGJT03} and scientific articles~\cite{DBLP:conf/uai/Rosen-ZviGSS04}. However, when processing short and ill-formed texts, such as app reviews and Twitter messages, the performance of LDA will be inevitably compromised~\cite{DBLP:conf/www/YanGLC13}. 

\textbf{Biterm Topic Model (BTM)}~\cite{DBLP:conf/www/YanGLC13} is designed specifically for modeling topics of short texts. BTM extends a document into a set of biterms (\textit{i.e.}, two terms) which includes all combinations of any two distinct words appearing in one document. In this way, BTM can enrich the short texts by explicitly modeling the word co-occurrence patterns. In BTM, instead of assuming that each word belongs to one topic, it assumes that each biterm relates to one topic. BTM has demonstrated better performance than LDA in modeling short texts~\cite{DBLP:conf/naacl/Sridhar15a,DBLP:journals/ese/ChenTH16}.

% with all combinations of any two distinct words appearing in the document. In another word, BTM enriches the information for the original short text by explicitly model the word co-occurrence pattern. BTM has a similar assumption to LDA; the only difference is, each biterm (word-pair) in the document belongs to one topic. BTM demonstrates better performance in modeling short text, such as tweets~\cite{DBLP:conf/naacl/Sridhar15a} and user reviews~\cite{DBLP:journals/ese/ChenTH16}.

\textbf{Joint Sentiment/Topic Model (JST)}~\cite{DBLP:conf/cikm/LinH09} is also a variant of LDA, but involves the sentiment of each topic. JST assumes that each word should be associated with one sentiment polarity, such as positive, neutral, and negative. The output of JST is also two matrices but with three dimensions: (1) Document-sentiment-topic matrix, denoted as $\Theta\in\mathbb{R}^{D\times3\times K}$ where $3$ is the number of sentiment polarities (positive, negative, and neutral); (2) Sentiment-topic-word matrix, \textit{i.e.}, $\Phi\in\mathbb{R}^{3\times K\times V}$.

% by considering the sentiment orientation of text. Sentiment polarities are intuitively dependent on topics or domains. For example, though the adjective ‘fast’ in a phrase such as ‘load fast’ may have a positive orientation in a video app review, it could also have a negative orientation in a phrase like ‘die fast’ in a review complaining about battery consumption. Therefore, jointly modeling topic and sentiment should serve a critical function in helping
% users in terms of opinion mining and issue detection. JST assumes that each word should be assigned with one of the sentiment polarities among positive, negative, and neutral. The output of JST is two $3$-dimensional matrix: (1) document-sentiment-topic matrix $\Theta\in\mathbb{R}^{D\times3\times K}$, (2) sentiment-topic-word matrix $\Phi\in\mathbb{R}^{3\times K\times V}$.

% by modeling an example of user review

\begin{figure}[t]
% \vspace{1mm}
    \centering
    \resizebox{\linewidth}{!}{
    \includegraphics{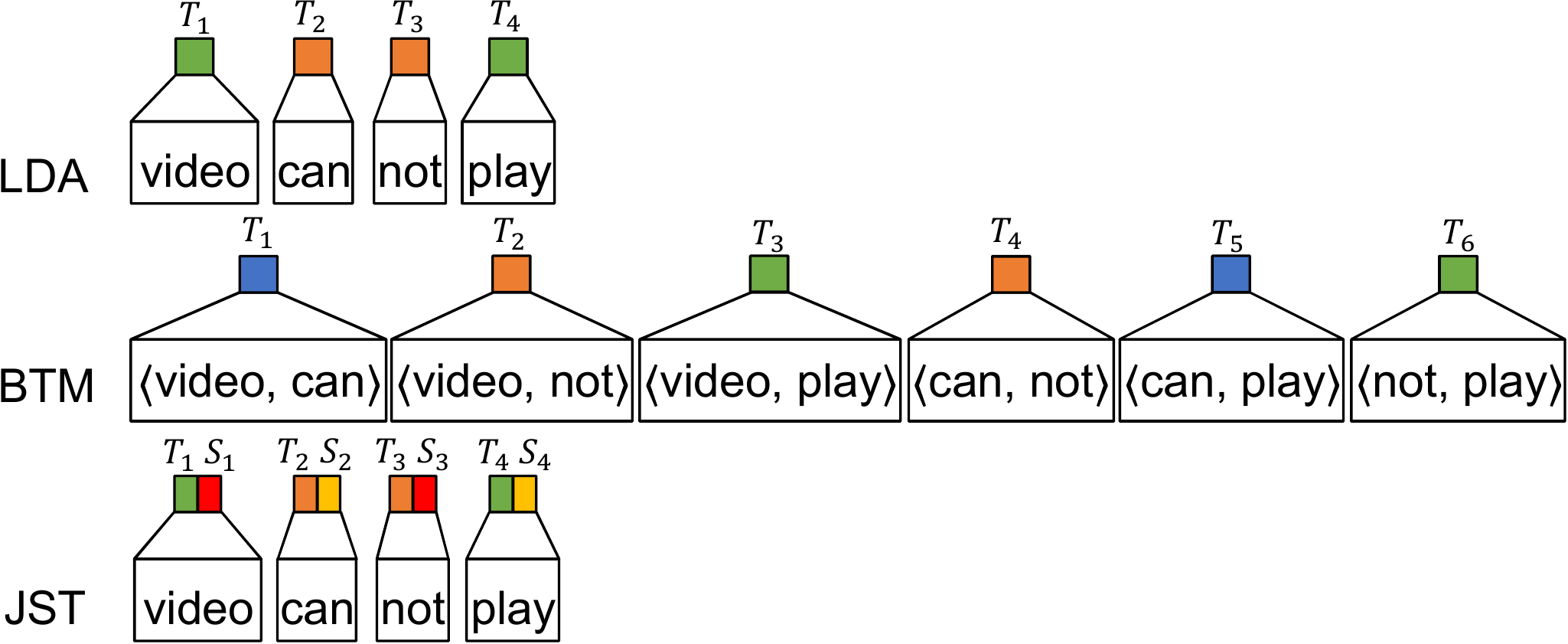}
    }
    \caption{Illustration of three typical topic modeling approaches, including LDA, BTM, and JST, based on an example of user review (``\textit{video can not play}''). The symbols $T_i$ and $S_i$ denote the inferred topic and sentiment of the $i$-th token in the review, respectively. \yun{Different colors indicate different topics or sentiments conveyed by the corresponding words.}
    }
    \label{fig:tms}
\end{figure}

Figure~\ref{fig:tms} illustrates the differences among the above topic models in terms of their input and output. However, documents mostly come as ephemeral streams in most scenarios, such as scientific articles and Twitter messages, and thus the topics and subordinate attributes (\textit{e.g.}, word distributions) in the documents are time-evolving~\cite{DBLP:conf/uai/AhmedX10}. To capture such topic variations, online topic models, including Online LDA (OLDA)~\cite{DBLP:conf/cikm/LinH09}, have been proposed. The output of online topic models is topic distributions along with consecutive time slices, and Table~\ref{tab:tm_output} can be regarded as the topic distributions of the documents in one time slice.

% several aspects of the latent structure such as the topics' distributions and popularity are time-evolving~\cite{DBLP:conf/uai/AhmedX10}. Online topic models allow the topic modeling techniques to work in an online fashion such that they incrementally build updated models when new documents appear~\cite{DBLP:conf/icdm/AlSumaitBD08}, which are more practical for analyzing continuously-updated data or text streams. Due to the real-time nature, online topic models have been extensively studied in the scenario of social media (\textit{e.g.}, Twitter) to detect newsworthy events. The output of the models is the topic distributions along with time slices, and Table~\ref{tab:tm_output} shows the result of one time slice. The meaning of each topic can be inferred by the top-ranked words based on their weights.

\begin{table}[]
	\centering
	\caption{Example for illustrating the output of topic models. The top-five words of each topic are listed, with corresponding probabilities. The meaning of each topic can be manually deduced from the top-ranked words.}
	\label{tab:tm_output}
	\includegraphics[width=0.42 \textwidth]{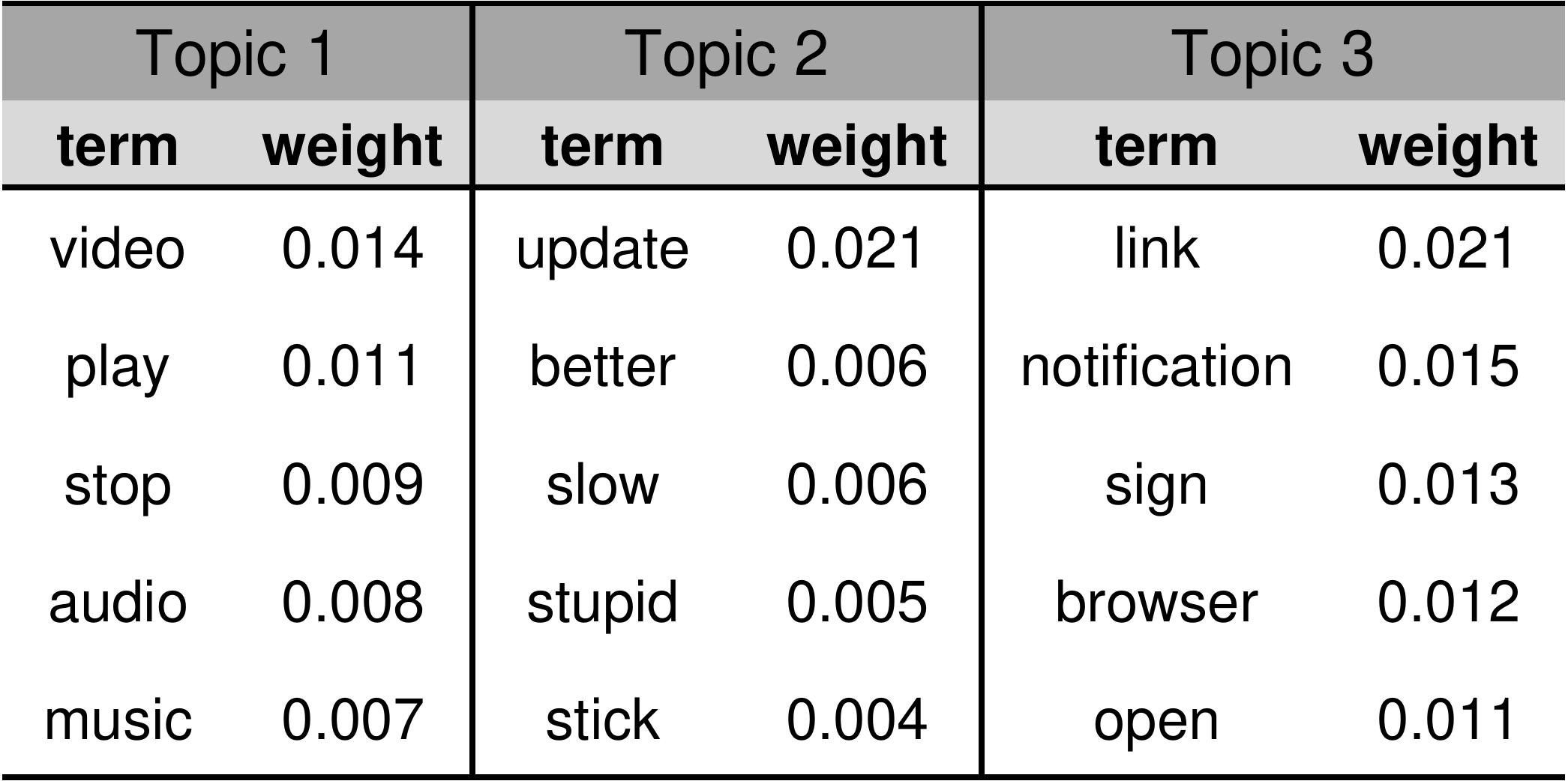}
\end{table}

\subsection{Word Embeddings}
Word embedding (also known as distributed representation~\cite{NIPS2013_5021,pennington2014glove}) is one of the most popular techniques that represent document vocabulary by training on a large text corpus. They map each word to a low-dimension real-valued vector and are capable of capturing the context of a word based on the semantic similarity relations with other words. The words that exhibit the same semantics have similar vector representations. For example, suppose the word ``\textit{photo}'' is represented as [0.53, -0.21, 0.02] and the word ``\textit{image}'' is represented as [0.49, -0.35, 0.01]. From their vectors, we can estimate their distance and identify their semantic relation. Word embedding is usually implemented through training a machine learning model such as CBOW and Skip-Gram~\cite{NIPS2013_5021} on large datasets. Phrases, sentences, and documents can also be embedded as vectors based on word embedding techniques. For example, a simple way of sentence embedding is to compute the average word embeddings in the sentence~\cite{DBLP:conf/icml/LeM14}.

% such as Word2vec and Glove\cite{pennington2014glove} 

\section{Overview of MERIT}\label{sec:overview}
\begin{figure}[t]
% \vspace{1mm}
    \centering
    \resizebox{0.95\linewidth}{!}{
    \includegraphics{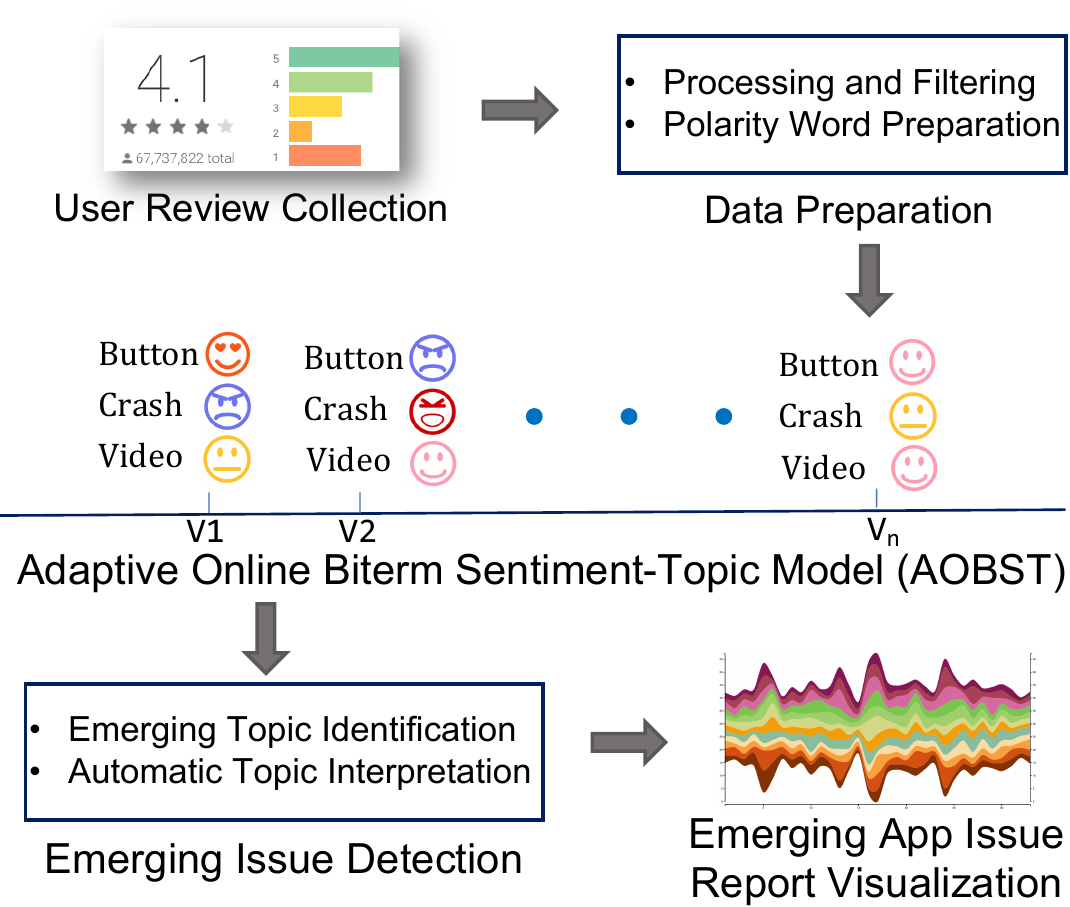}
    }
    \caption{Overview of the proposed framework - MERIT.
    }
    \label{fig:overall_framework}
\end{figure}

Figure~\ref{fig:overall_framework} illustrates the detailed steps of the proposed framework - MERIT, mainly including five steps: user review collection, data preparation (Section~\ref{ssec:process}), training and use of an adaptive online biterm sentiment-topic model (Section~\ref{sec:approach}), emerging issue detection (Section~\ref{sec:emerging}), and emerging app issue report visualization (Section~\ref{sec:real_time_analysis}). 
\subsection{Data Preparation}\label{ssec:process}
Since user reviews are usually written on mobile phones with limited keyboards on mobile screens, they often contain a large number of noisy words, such as misspelled words and abbreviations (\textit{e.g.}, ``\textit{asap}'').
% and \yun{informal words (\textit{e.g.}, ``\textit{asap}'')}. 
In the following, we elaborate on the preprocessing method and also the method to prepare {\em polarity-carrying words} (\textit{i.e.}, the words that carry sentiment polarities, \textit{e.g.}, positive or negative) for the subsequent modeling process.

\subsubsection{Preprocessing and Filtering}\label{sssub:preprocess}
% Since we adopt the preprocessing method in~\cite{gao2018online}, 
We adopt the preprocessing method described in~\cite{gao2018online}. For completeness sake, we briefly describe the steps here. We first convert all the words into their lowercase and then lemmatize them into their root forms following the lemmatization method described in~\cite{DBLP:conf/kbse/VuNPN15}. We also adopt the rule-based methods in~\cite{DBLP:conf/kbse/VuNPN15,DBLP:conf/issre/ManGLJ16} to rectify repetitive words (\textit{e.g.}, ``\textit{very very good}'' to ``\textit{very good}''), misspelled words, and remove non-English words. Then, we extract phrases (mainly referring to two consecutive words following our previous work~\cite{gao2018online}) for the topic interpretation procedure in Section~\ref{sec:emerging}. We use PMI (Point-wise Mutual Information)~\cite{DBLP:journals/coling/ChurchH90}, a measure of word association in information theory and statistics, to identify meaningful phrases based on co-occurrence frequencies.
\begin{equation}\label{equ:pmi}
    PMI(w_i, w_j) = \log\frac{p(w_iw_j)}{p(w_i)p(w_j)},
\end{equation}

\noindent where $p(w_iw_j)$ and $p(w_i)$ (or $p(w_j)$) indicate the co-occurrence probability of the phrase $w_iw_j$ and the probability of the word $w_i$ (or $w_j$) in the review collection. A higher PMI value illustrates that the two words appear together more frequently, and are more likely to be a meaningful phrase. The phrases with PMIs larger than a manually-defined threshold\footnote{The threshold is experimentally set.} are extracted. Finally, we reduce the non-informative words using the predefined list of to-be-filtered words proposed by Gao et al.~\cite{gao2018online}, including abbreviations (\textit{e.g.}, ``\textit{ur}'') and stop words (\textit{e.g.}, ``\textit{is}'').

\subsubsection{Polarity Word Preparation}
To infer the sentiment affiliated with each topic, we first create a list containing words and their corresponding polarities. We \yun{build}
% built
the word list leveraging opinion lexicons published by Hu and Liu~\cite{DBLP:conf/kdd/HuL04,sentimentword}, which include 2,006 positive words and 4,783 negative words identified from customer reviews. \yun{However,} since the published lexicons are from product reviews, there may exist discrepancies with the app review scenario. To mitigate the discrepancy, we \yun{adopt the collected reviews and} extract 15,704 opinion words, including verbs, adverbs, or adjectives based on part-of-speech tagging~\cite{postagging}. \yun{Due to the huge effort in manually labeling polarities of all the extracted opinion words, we randomly select 500/15,704 words based on their frequencies for manual labeling. The selected words are a statistically significant proportion of the whole opinion words, providing us with a confidence level of 95\% and a confidence interval of 5\%.}
% from the whole collected reviews and manually label polarities of the randomly selected 500/15,704 words providing us with a confidence level of 95\% and a confidence interval of 5\%. 
The labeling process is conducted by the first author and two Computer Science Ph.D. students. Each word needs to be labeled by two of the annotators, and the label options can be ``1 (positive)'', ``0 (neutral)'', or ``-1 (negative)''. The labeling achieves 0.79 agreement rate\footnote{The agreement rate is computed based on Cohen's kappa~\cite{cohen1968weighted}.} and full agreement after discussion. Table~\ref{tab:word_polarity} lists some examples of the labeled word polarities. We combine the manually-labeled \yun{500} opinion lexicons \yun{from the collected app reviews} with the published ones~\cite{DBLP:conf/kdd/HuL04} as our \textit{word polarity list}\footnote{\yun{For the coincident words in the 500 opinion lexicons and 6,789 published polarity words, we choose their polarities as the labels in app review scenario. In total, we obtain 7,215 opinion words and their polarities.} Full list of the word polarities can be found in our replication package.}. \yun{By integrating the opinion words from app reviews, we can mitigate the polarity discrepancy caused by solely using the polarity words from product reviews.}

\begin{table}[ht]
	\small
% 	\captionsetup{aboveskip=-1pt}
	\caption{Examples of labeled word polarities. Positive, neutral, and negative sentiments are indicated as ``1'', ``0'', and ``-1'', respectively.}
	\label{tab:word_polarity}
	\center
	\scalebox{1}{%
	\begin{tabular}{|l|r||l|r|}
		%       \rowcolor{gray!50}
		\hline
		\textbf{Word} & \textbf{Sentiment} & \textbf{Word} & \textbf{Sentiment} \\
		\hline
		\hline
		comfortable & 1 & unnecessary & -1 \\
		\hline
		buggy & -1 & learn & 0 \\
		\hline
		weird & -1 & unclear & -1 \\
		\hline
		beneficial & 1 & exclude & -1 \\
		\hline
		consistent & 1 & blame & -1 \\
		\hline
		inform & 0 & unlock & 0 \\
	    \hline
	\end{tabular}
}
\end{table}

All the non-filtered words, phrases extracted following Equ.~(\ref{equ:pmi}) (where the words in each phrase are concatenated with ``\_''), and the word polarity list are fed into the topic modeling process.

% of verbs, adverbs, or adjectives

\subsection{Adaptive Online Biterm Sentiment-Topic Model (AOBST)}\label{sec:approach}

\begin{figure*}[t]
% \vspace{1mm}
    \centering
    \resizebox{0.85\linewidth}{!}{
    \includegraphics{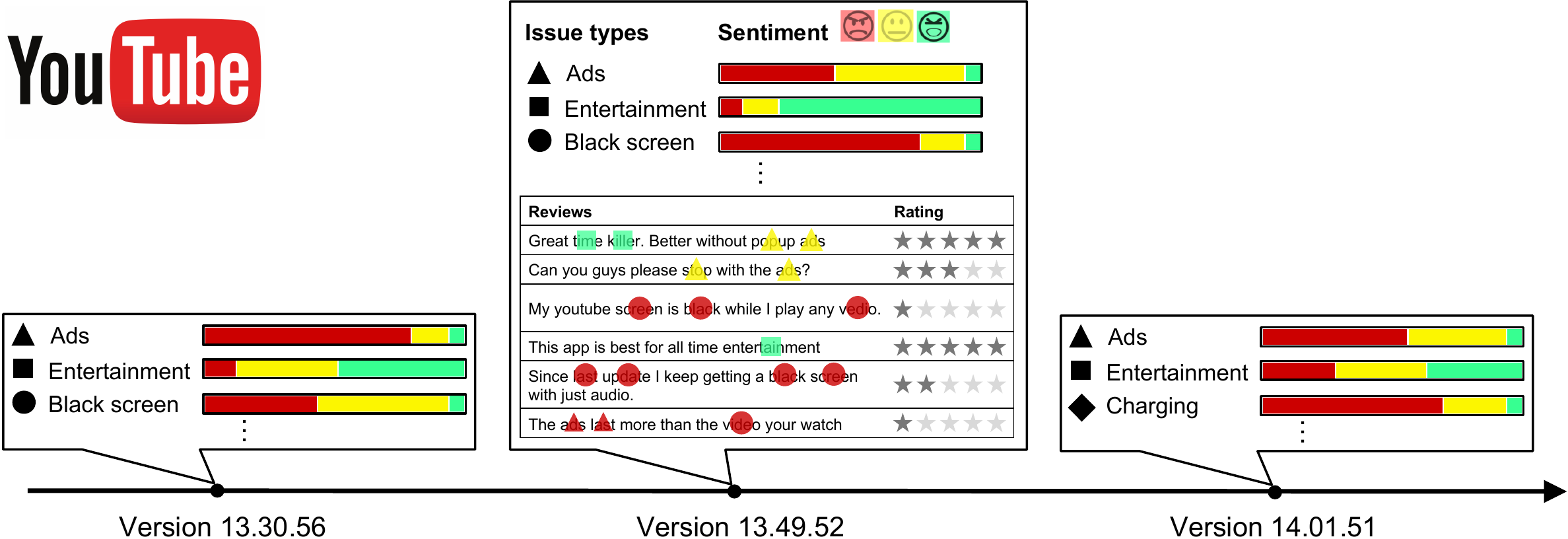}
    }
    \caption{Illustration of the output of the proposed adaptive online biterm sentiment-topic model (AOBST). The horizontal axis represents examples of released consecutive versions for YouTube on App Store. For each version, the AOBST model generates its topic distributions, indicated by different shapes, and the corresponding sentiment distributions (displayed with color bars beside the topics). Different colors represent different sentiment levels, from \colorbox{red}{\makebox(18,5){\strut angry}} to \colorbox{yellow}{\makebox(33,5){just-so-so}} to \colorbox{green}{\makebox(20,5){happy}}. 
    % The exclamation marks \textcolor{red}{!!!} beside an issue type (\textit{e.g.}, in Version 14.01.51) indicate a significant change in the topic.
    % a color bar (\textit{e.g.}, in Version 13.49.52) or
    }
    \label{fig:aojstb}
\end{figure*}

Inspired by existing topic modeling techniques~\cite{DBLP:conf/cikm/LinH09,DBLP:conf/www/YanGLC13}, we propose a novel unsupervised model named AOBST (Adaptive Online Biterm Sentiment-Topic Model) for jointly modeling the topics and sentiment of app reviews. We will first illustrate the proposed biterm sentiment-topic model for building connections between topics and sentiment, and then elaborate on its online adaption.

\subsubsection{Biterm Sentiment-Topic Model}\label{sssec:bst}

\begin{figure*}[ht]
    \centering
    \begin{subfigure}[b]{0.3\textwidth}
        \includegraphics[width=\textwidth]{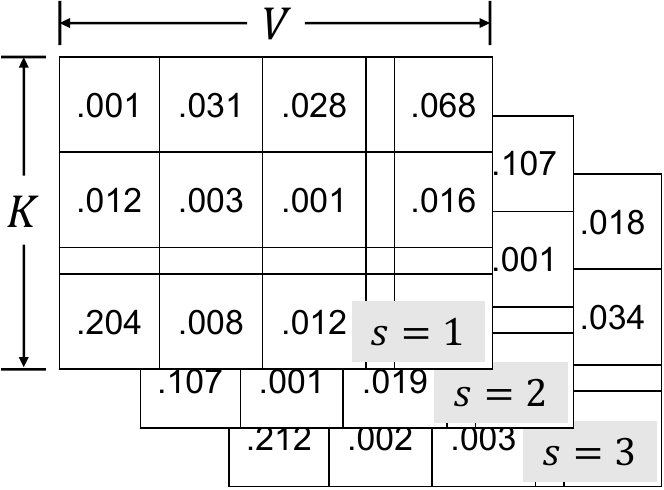}
        \caption{Example of the 3-D matrix $\Phi$.}
        \label{fig:phi_1}
      \end{subfigure}
      \hfill
      \begin{subfigure}[b]{0.65\textwidth}
        \includegraphics[width=\textwidth]{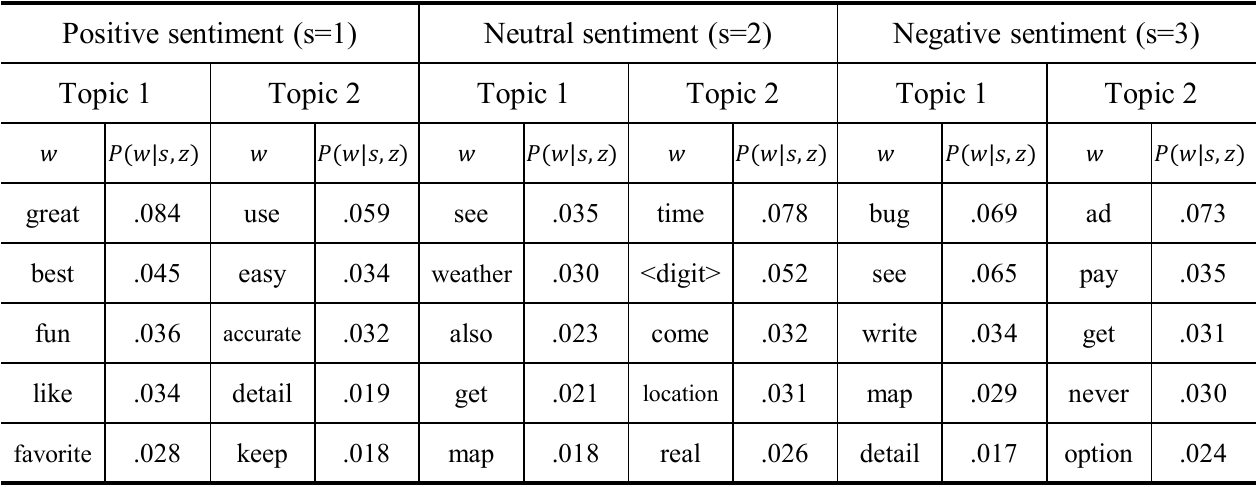}
        \caption{Example of the output of BST, with top-five words listed for each topic.}
        \label{fig:phi_2}
      \end{subfigure}
    \caption{Illustration of the learned sentiment-topic-word matrix $\Phi$ from BST.}
	\label{fig:btm_sample}
\end{figure*}

% assumption, sampling process
To address the first two limitations described in Section~\ref{sec:introduction}, including the short-length and sentiment characteristics of app reviews, we propose a Biterm Sentiment-Topic (BST) model. The BST model is built upon BTM and JST, since BTM has shown better performance than LDA in modeling short texts and JST can jointly model topics and sentiment. We introduce the details of the proposed BST below.

% Due to the short and noisy nature of app reviews, biterm topic model should perform better than traditional topic models, such as LDA and PLSA. Besides, app issues tend to be negatively expressed by users via reviews so that sentiment-aware topic modeling could detect issue-oriented topics more accurately than the models solely involved topics. Therefore, we propose the biterm sentiment-topic model (BST) to jointly model the sentiment and topics, and address the first two limitations described in Section~\ref{sec:introduction}.
% BST is based on a fully generative model that describes how an app review is created.

BST assumes that each app review is a set of biterms $B$, and each biterm $b=(w_i, w_j)$ belongs to one sentiment polarity $s$ and one topic $z$. The modeling process can be described as below:

% Intuitively, BST assumes that each app review is an aggregation of biterms (word pairs), for each biterm, it is assigned with a sentiment label and a sentiment-aware topic. The specific generative process of app review in BTM can be described as follows:
\begin{itemize}
\item Construct a sentiment distribution $\pi\sim Dir(\gamma)$.
\item For each sentiment polarity $s$:
\begin{itemize}
\item Construct a topic distribution for sentiment $s$, $\theta_s\sim Dir(\alpha)$.
\item For each topic $z$:
\begin{itemize}
    \item Construct a word distribution for sentiment $s$ and topic $z$, $\phi_{s,z} \sim Dir(\beta)$. 
\end{itemize}
\end{itemize}
\item For each biterm $b$ in the biterm set $B$:
\begin{itemize}
\item Choose a sentiment polarity $s_b\sim Multi(\pi)$.
\item Choose a topic assignment $z_b\sim Multi(\theta_{s_b})$.
\item For each word $w_i$ in the biterm
\begin{itemize}
    \item Choose a word $w_i$ based on the distribution over words, \textit{i.e.}, $w_i\sim Multi(\phi_{s_b, z_b})$, where $z_b$ and $s_b$ denotes the topic and sentiment polarity, respectively.
\end{itemize}
\end{itemize}
\end{itemize}

The hyperparameters $\gamma$, $\alpha$, and $\beta$ in BST can be treated as the prior counts of the sampled sentiment polarity $s$, the sampled topic $z$ associated with sentiment polarity $s$, and the sampled words for topic $z$ and sentiment polarity $s$, respectively. $Dir(\cdot)$ and $Multi(\cdot)$ represent Dirichlet distribution and multinomial distribution parameterized by $\cdot$, respectively. The probability of a biterm $b=(w_i, w_j)$ can be calculated as:

% times of topic $j$ associated with sentiment label $s$ is sampled respectively, before having observed any actual words. Similarly, $\beta$ can be interpreted as the prior observation number of times words sampled from topic $j$ associated with sentiment label $s$.
% We can manipulate the hyperparameter to make BST initialized with the previous observation and instruct the modeling for the current data, which is the idea of online BST learning, and we will introduce it in the next section.
% probability
% The joint probability of a biterm $b=(w_i, w_j)$ can be written as:
\begin{equation}
\begin{aligned}
    P(b)&=\sum_{s,z}P(z|s)P(w_i|s,z)P(w_j|s,z) \\
    &=\sum_{s,z}\theta_s \phi_{i|s, z} \phi_{j|s, z}
\end{aligned}
\end{equation}
% Thus the likelihood of the whole corpus is:
% \begin{equation}
% \begin{aligned}
%     P(B)&=\prod_{i,j}\sum_{s,z}\theta_s \phi_{i|s, z} \phi_{j|s, z}
% \end{aligned}
% \end{equation}

The parameter matrices, \textit{i.e.}, $\{\Theta\in \mathbb{R}^{3\times K}, \Phi\in\mathbb{R}^{3\times K\times V}\}$, of BST can be inferred through Gibbs sampling~\cite{steyvers2007probabilistic} efficiently, given all the biterms $B$.
% outputs
% , which can be constructed from each review by the pairwise combination of words
The parameter matrix $\Phi$ is the sentiment-topic-word matrix, with an example shown in Figure~\ref{fig:phi_1}. The first dimension of $\Phi$ is the sentiment polarity (\textit{i.e.}, $s\in \{1, 2, 3\}$ for representing each of the three sentiment --- 1 = negative, 2 = neutral, 3 = positive). We can regard the second and third dimensions of $\Phi$ as a topic-word matrix ($\mathbb{R}^{K\times V}$), with each row indicating the probability distribution over words for the topic. By inspecting the topic examples extracted from $\Phi$, shown in Figure~\ref{fig:phi_2}, we can discover that the topics exhibit different sentiment polarities from the sentiment perspective.

% three sentiment polarities in total,
% it is evident that topics under positive sentiment label and topics under negative sentiment label indeed have words in different sentiment polarities.

\subsubsection{Adaptive Online Joint Sentiment-Topic Tracing}
In the previous section, we have introduced BST for inferring sentiment-aware topics from an app review collection. In this section, we will describe an online adaption of BST to trace topic variations of review collections from consecutive app versions. 

% To automatically infer the sentiment associated with topics, we exploit the prepared polarity words as prior knowledge and detect sentiment and topics simultaneously from text. The details are described below.

We first divide collected app reviews according to app versions, denoted as $R=\{R_1,R_2,...,R_t,...\}$, where $R_t$ indicates all the reviews pertaining to the $t$-th app version. In order to capture the topic evolution along with versions, we apply an adaptively online topic modeling mechanism~\cite{gao2018online} to BST, \textit{i.e.}, adaptive online biterm sentiment-topic model (AOBST). AOBST adaptively connects the sentiment-topic word distributions in
% sentiment-topic distributions
previous app versions with the prior for the word distribution $\beta$ of current app version.
Specifically, we denote the sentiment-topic word distributions in previous $\omega$ version as $\{\phi^{t-1},...,\phi^{t-i},...,\phi^{t-\omega}\}$, where $\omega$ is the version window size determining the number of previous versions to be considered for analyzing the sentiment-topic word distributions of the current version. The \textit{connection strength} $\eta$ between the sentiment-topic word distribution $\phi^{t-i}$ in the previous $i$-th version and the prior $\beta^t$ of the current $t$-th version is defined as their similarity, which is calculated in the following:

\begin{equation}
\eta_{s,z}^{t,i}=\frac{\exp (\phi_{s,z}^{t-i}\cdot \beta_{s,z}^{t-1})}{\sum_{j=1}^\omega\exp (\phi_{s,z}^{t-j}\cdot \beta_{s,z}^{t-1})},
\end{equation}

\noindent where $i$ denotes the $i$-th previous version ($1\le i\le \omega$), and $s$ and $z$ indicate the current sentiment and topic respectively. The dot product $\phi_{s,z}^{t-i}\cdot \beta_{s,z}^{t-1}$ computes the similarity between the word distribution of the previous $i$-th version $\phi_{s,z}^{t-i}$ and the prior of $(t-1)$-th version $\beta_{s,z}^{t-1}$. Such adaptive connection can endow the sentiments and topics of the previous versions with different contributions to the sentiment-topic inference of the current version~\cite{gao2018online}. The prior $\beta^t$ is calculated as:

\begin{equation}
\beta_{s,z}^t = \sum_{i=1}^{\omega}\eta_{s,z}^{t,i}\phi_{s,z}^{t-i}.
\end{equation}

Based on AOBST, we can trace the variations of topics for different sentiment polarities along with app versions, as shown in Figure~\ref{fig:aojstb}. We describe the approaches to detect the emerging topics and automatically interpret the topic meanings with phrases and sentences in the next section. Since we aim at detecting app issues, which are generally expressed in an unfavorable manner by users, we focus on the negative topics during emerging issue detection.

\subsection{Emerging Issue Detection}\label{sec:emerging}
In this section, we describe how we determine the emerging app issues based on the evolution of the topics belonging to negative sentiment along with app versions.

\subsubsection{Emerging Topic Identification}
% Since negative topics are generally more concerned by users and thereby crucial for developers, we propose to identify anomaly topics from those who are inferred as negative by AOBST. 
Following the previous study in anomaly detection~\cite{DBLP:journals/csur/ChandolaBK09}, anomalies are defined as data points that deviate significantly from the majorities within a group. In this work, we define the emerging topics as those present obvious differences with the counterparts in the previous versions. The identified topics are regarded as emerging topics. We focus on the topics inferred as negative during emerging topic detection.

We compute the difference of the $z$-th negative topics between two consecutive versions, \textit{e.g.}, $\phi_z^t$ and $\phi_z^{t-1}$ and adopting the classic Jesen-Shannon (JS) divergence~\cite{DBLP:journals/tit/Lin91}. Higher JS value indicates that the two topic distributions exhibit a larger difference. In this way, we generate a $\omega\times K$ divergence matrix $D_{JS}$ (where $\omega$ and $K$ are the number of window size and topics respectively) for the versions in a window. We then use the typical outlier detection method~\cite{DBLP:journals/widm/RousseeuwH11} to detect the anomalies:

\begin{equation}
    \frac{\{D_{JS}\}_z^t-\overline{D_{JS}}}{\sigma} > \delta,
\end{equation}

\noindent where $\overline{D_{JS}}$ and $\sigma$ denote the mean and standard deviation of all the values in the computed $D_{JS}$ matrix. The threshold $\delta$ determines how far the current JS divergence differs from the expected divergence value as compared to the typical difference (\textit{i.e.}, the standard deviation). We set $\delta=1.25$ for accepting 10\% of the total topics as anomaly topics following our previous work~\cite{gao2018online}.

% Besides, we compute divergences $D_{JS}$  in a window size $\omega$. It means that we only compare the topics in current versions with those in previous $\omega$ versions during divergence computation. 

\subsubsection{Automatic Topic Interpretation}\label{sssec:topic_inter}
% \subsection{Adding word embedding}
By directly observing the top few words per topic as shown in Table~\ref{tab:tm_output}, developers may find it difficult to capture the concrete meaning of each topic. In this section, we aim at automatically interpreting each topic. We choose phrases and sentences for the interpretation, since the meanings of single words may be ambiguous and entire reviews with more than one sentence can express totally different aspects. The phrases are prioritized from the candidates extracted during the preprocessing step in Section~\ref{sssub:preprocess}. To solve the third limitation described in Section~\ref{sec:introduction}, \textit{i.e.}, ineffectiveness of the topic labeling approach in~\cite{gao2018online}, we combine word embeddings with topic distributions as the semantic representations of words. We denote the proposed New Topic Labeling approach as NTL. The details are described as follows.

\vspace{0.2cm}
\noindent\textbf{(1) Interpreting Topics with Phrases.}

% because word embeddings have been proven to be more effective in semantic representation than topic-based strategies~\cite{DBLP:conf/coling/JiangSLW16,DBLP:conf/acl/HuT16,DBLP:conf/acl/LiCZM16}. 

The similarity $Score$ between each phrase candidate $a$ and topic $\phi_z^t$ is calculated in two levels: \textbf{topic level} and \textbf{embedding level}.

\textbf{Topic Level.} The topic distributions over words obtained from AOBST indicate the topical relevance of each word in the vocabulary to the topic. If one phrase candidate and topical words are closer to each other in the topic space, the candidate is more representative of that topic. We employ the method in~\cite{DBLP:conf/kdd/MeiSZ07} to measure the topical similarity between the phrase candidate $a$ and target topic $\phi_z^t$, defined as:

\begin{equation}
    Sim_{topic}(a, \phi_z^t) = -D_{KL}(a||\phi_z^t),
\end{equation}

\noindent where $D_{KL}$ denotes the Kullback-Leibler (KL) divergence~\cite{DBLP:journals/tit/Lin91} which is utilized to measure the distance between two probabilistic vectors.

\textbf{Embedding level.} In the embedding space, if the phrase candidates and topical words are closer to each other, the candidates are more semantically representative. For this, we propose a semantic match score based on the attention mechanism~\cite{journals/corr/ParikhT0U16}:

\begin{equation}
Sim_{embed}(a, \phi_z^t) = \sum_{w} \frac{\exp(e_{a, w})}{\sum_{w} \exp(e_{a, w})}\phi_{z,w}^t,
\end{equation}

\noindent where $e_{a,w_i}$ and $\exp(\cdot)$ indicates the cosine similarity score between two embeddings and its exponential format. The fractional term represents the similarity match score between the phrase candidate and topical words in the embedding space. A phrase candidate with a higher match score with the top topical words will be ranked higher.

\vspace{0.2cm}
\noindent\textbf{(2) Interpreting Topics with Sentences.}

For a sentence candidate $s$, its topic-level and embedding-level similarity scores are computed as below.

\textbf{Topic level.} A sentence candidate is more representative of one topic if it comprises more words presenting higher topic relevancy to that topic. The similarity between a sentence candidate $s$ and topic $\phi_z^t$ is computed as:

\begin{align}
    Sim_{topic}(s, \phi_z^t) & = -D_{KL}(s||\phi_z^t) \nonumber\\
    & \approx \sum_w -D_{KL}{(w||\phi_z^t)}p(w|s),
\end{align}
\noindent where $p(w|s)$ denotes the term frequency of $w$ in the sentence $s$. 

\textbf{Embedding Level.} Similarly, we calculate the embedding-level similarity of one sentence to the topic based on its constituent words, defined as:

\begin{equation}
    Sim_{embed}(s, \phi_z^t) = \sum_w Sim_{embed}(w, \phi_z^t).
\end{equation}

The overall similarity score of each candidate $l$ (indicating a phrase $a$ or sentence $s$) is determined based on the combination of both topic-level and embedding-level scores:

\begin{equation}\label{equ:mu}
    Score(l, \phi_z^t) = Sim(l, \phi_z^t) - \frac{\mu}{K-1}\sum_{j\neq z} Sim(l,\phi_j^t),
\end{equation}

\noindent and

\begin{equation}\label{equ:labeling}
    Sim(l, \phi_z^t) = m*Sim_{topic}(l, \phi_z^t) + (1-m)*Sim_{embed}(l, \phi_z^t),
\end{equation}

\noindent where $m\in (0,1)$ is a real-valued weight for balancing the two levels of similarity scores, $l$ can be a phrase candidate $a$ or sentence candidate $s$, and $\mu$ is a penalty factor to adjust the similarities to other topics.

\subsection{Emerging App Issue Report Visualization}\label{sec:real_time_analysis}
\begin{figure*}[h]
% \vspace{1mm}
    \centering
    \begin{subfigure}{0.75\textwidth}
    \centering
    \includegraphics[width=\textwidth]{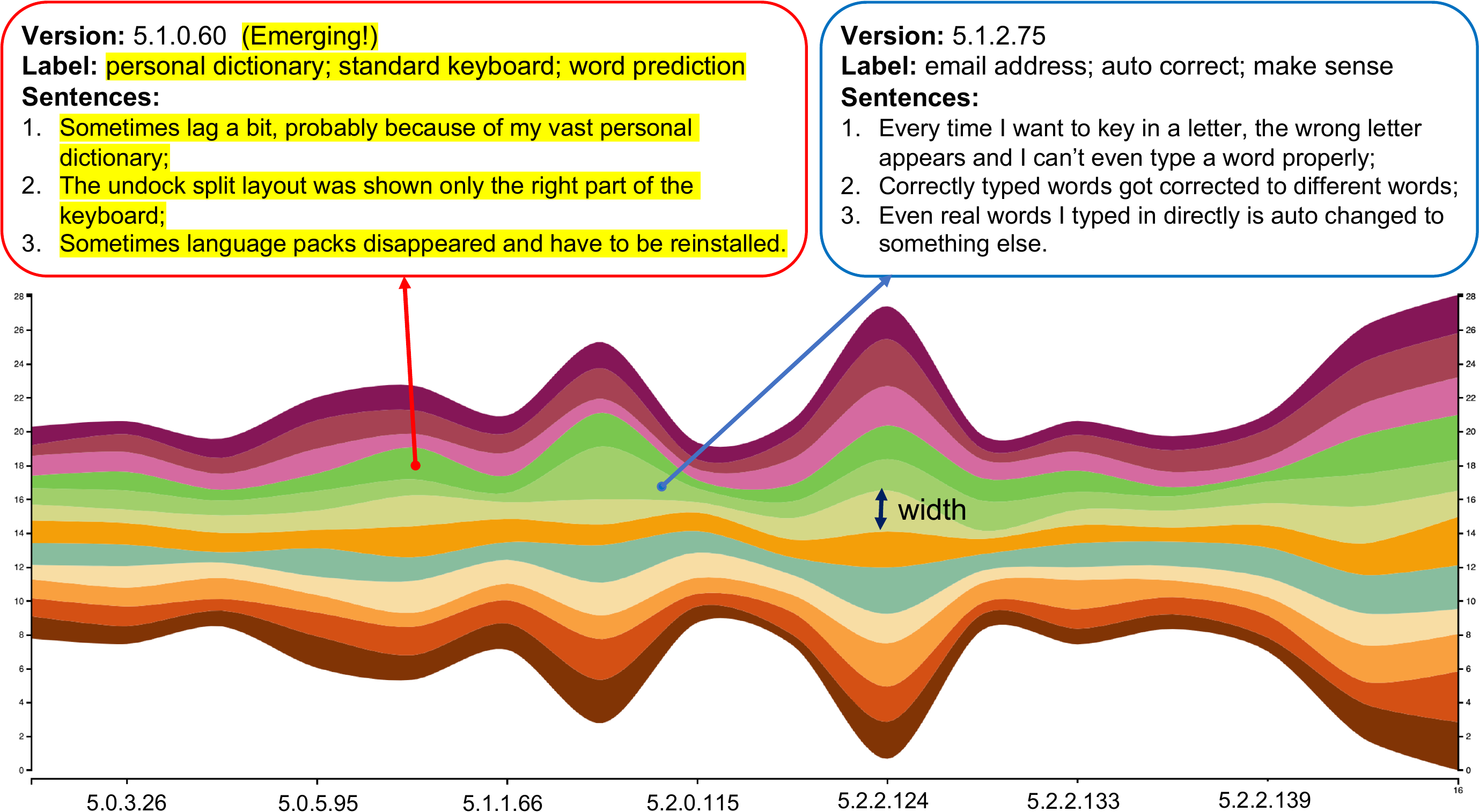}
    \end{subfigure}
    \hfill
    \begin{subfigure}{0.23\textwidth}
    \centering
    \includegraphics[width=\textwidth]{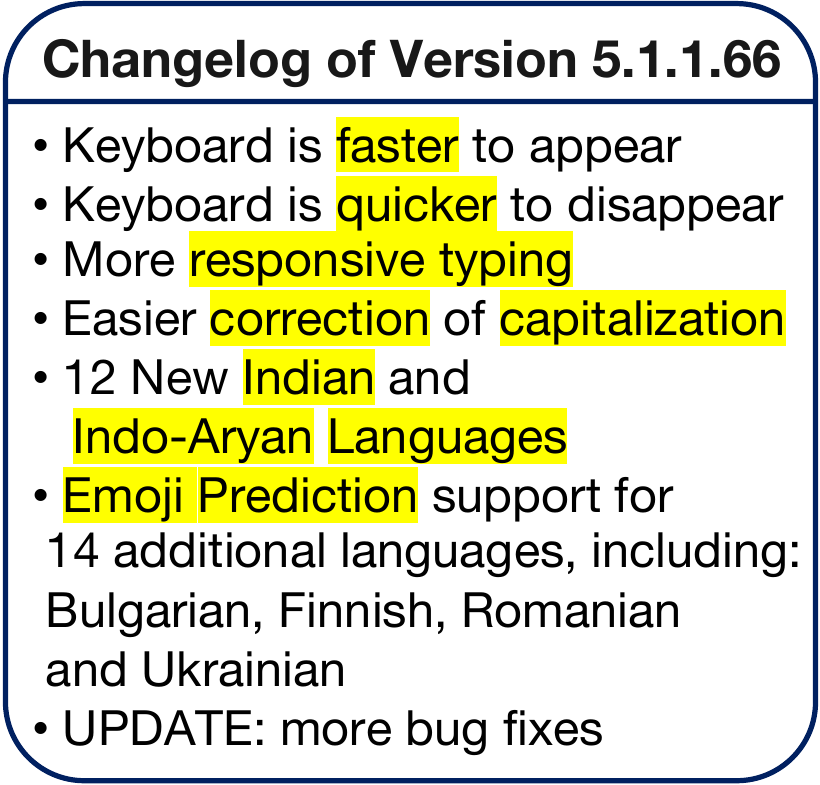}
    \end{subfigure}

    \caption{Issue river of the SwiftKey app (left) and changelog of the version 5.1.1.66 (right). For the issue river, the whole topic flow is visualized as a river, and the number of topics $K$ is set as 12, corresponding to 12 branches of the river. The horizontal axis presents consecutive app versions, and the branches with larger widths illustrate that the corresponding issues relatively concern users more for those versions. For the changelog, we highlight keywords in yellow background.}
    \label{fig:river}
\end{figure*}

For facilitating developers to efficiently understand the identified emerging app issues, we visualize the evolution of app issues along with versions based on \textit{issue river}~\cite{gao2018online}. Figure~\ref{fig:river} (Left) shows an example for Swiftkey for Android. The whole river represents all the app issues, and different branches indicate different topics. The \textit{width} of each branch $k$ presents the user-concern degree of the issue for the corresponding version $t$, defined as:

\begin{equation}
    width_{k}^{t} = \sum_{a}\log Count(a)* \phi_{k}^{t},
\end{equation}

\noindent where $Count(a)$ means the count of the phrase label $a$ in the review collection of the $t$-th version. So, wider branches are of more concern to users. By moving the mouse over one topic (\textit{i.e.}, branch), developers can track detailed issues along with versions, where the emerging ones are highlighted with yellow background, as shown on the top left box in Figure~\ref{fig:river}. We also show an example of changelog on the right of Figure~\ref{fig:river}. We can discover that the identified emerging issue \textit{lag during word prediction} was fixed by the next immediate version, described as ``\textit{More responsive typing}'' (the third item) in the corresponding changelog.

% present interactive and detailed report, including issue-based review retrieval and visualization.

% \subsubsection{Review Retrieval}
% We provide related reviews to developers when they are interested in specific app issues and want to look at more details. We treat the top words belonging to a topic as query $q$, and retrieve similar reviews in the topic space. For each review, its similarity score with the top words is computed as:

% \begin{equation}
%     Score(d) = \cos{(\frac{\sum_{w\in d} \phi_{w}}{len(d)}, \frac{\sum_{w\in q} \phi_{w}}{len(q)})}.
% \end{equation}

% We prioritize the related reviews to developers for their reference, as shown in Figure \yun{Add figure}.

% \subsubsection{Visualization}

\section{Experimental Setup}\label{sec:setup}
\subsection{Dataset}
We employ the same dataset by Gao et al.~\cite{gao2018online} for evaluation. Details of the dataset are shown in Table~\ref{tab:dataset}. The dataset includes 164,026 reviews (from August 2016 to April 2017) for six apps, from 89 versions in total. The apps are distributed in different categories, with two of them from the App Store and the others from Google Play.

\begin{table}[ht]
	\small
% 	\captionsetup{aboveskip=-1pt}
	\caption{Subject apps.}
	\label{tab:dataset}
	\center
	\scalebox{0.8}{%
	\begin{tabular}{|l|l|l|r|r|}
		%       \rowcolor{gray!50}
		\hline
		App Name & Category & Platform & \#Reviews & \#Versions \\
		\thickhline
		NOAA Radar & Weather & App Store & 8,363 & 16 \\
		\hline
% 		Waze & Navigation & App Store & 3,849& 9\\
% 		\hline
		YouTube & Multimedia & App Store & 37,718 & 33 \\
		\hline
% 		Snapchat & Photo \& Video & App Store & 23,073& 11\\
% 		\hline
		Viber & Communication & Google Play & 17,126 & 8 \\
		\hline
		Clean Master & Tools & Google Play & 44,327 & 7\\
		\hline
	    Ebay & Shopping & Google Play & 35,483 & 9 \\
	    \hline
	    Swiftkey & Productivity & Google Play & 21,009 & 16 \\
	    \hline
	   % Waze & Navigation & Google Play &9,601 & 12\\
	   % \hline
% 	    SHAREit & Tools & Google Play & 45,316 & 40\\
% 		\hline
	\end{tabular}
}
\end{table}

% In Table~\ref{tab:dataset}, we list the subject apps with the app name, category, platform, the number of reviews crawled, and the number of versions in the review collection. Overall, we obtain 164,026 reviews (from August 2016 to April 2017) for the six apps, from 89 versions in total. The apps are distributed in different categories, with two of them from App Store and the others from Google Play. With multiple categories and platforms, the generalization of MERIT can be ensured.

\subsection{Evaluation Methods}
% The app changelogs, \textit{i.e.}, our ground truth, are collected from App Annie. Since the prioritized issues of MERIT are in phrases and sentences, we manually extract key terms from these changelogs for verification. One example is illustrated in Table~\ref{tab:case_c}, with the key terms highlighted. For each key term in changelogs, we validate whether the term is covered by the prioritized issues. Since the \textit{word2vec} model~\cite{mikolov2013distributed} can accurately capture the semantic meanings of input terms based on their vector representations, we obtain the cosine similarities between each key term and the phrase-level issues based on the model. The key term is considered covered if its similarity to one of the issues is larger than 0.6~\cite{islam2008semantic}. For sentence-level issues, we split the sentences into terms (including phrases and words) and verify whether the key term in changelogs can be covered in a similar way. We employ such semi-automatic evaluation method to facilitate parameter adjustment and comparison with other methods. 
We use the keywords in changelogs as ground truth (one example shown on the right of Figure~\ref{fig:river}) and employ the three metrics as used by Gao et al.~\cite{gao2018online} for verifying the effectiveness of MERIT. We define an app issue to be successfully identified by MERIT and its baselines if its corresponding description in \yun{the} changelog \yun{of the immediate version} present a high similarity\footnote{The similarity is measured as the cosine score between the two vector representations in the word embedding space, and it is high if the cosine score is larger than 0.6~\cite{DBLP:journals/tkdd/IslamI08}.} with the identified issues (either at phrase level or sentence level). \yun{The evaluation method is based on the hypothesis that emerging issues need to be quickly solved in an updated version and thus are typically reflected in the changelog of the immediate version.} Here we use three performance metrics as used by Gao et al.~\cite{gao2018online} for verifying the effectiveness of MERIT. The first metric is for measuring the accuracy in detecting emerging issues, defined as $Precision_E$. The second is to evaluate whether our prioritized app issues (including both emerging and non-emerging issues) reflect the changes mentioned in the changelogs, defined as $Recall_L$. The last metric $F_{hybrid}$ is for measuring the balance between $Precision_E$ and $Recall_L$. Higher values of $F_{hybrid}$ indicate that changelogs are more precisely covered by detected emerging issues, and more changelogs are reflected in the prioritized issues.

% manually label the keywords in changelogs as ground truth (one example shown on the right of Figure~\ref{fig:river}) and \chuan{employ word-level cosine similarity in word embedding space to judge the matching of the predicted issues and ground truth (with similarity $>$ 0.6~\cite{DBLP:journals/tkdd/IslamI08}) in the next version changelogs}. 

\begin{equation}\label{eq:pnr}
\begin{split}
Prec&ision_{E} = \frac{I({E\cap G})}{I({E})},\ \ \  Recall_L=\frac{I({L\cap G})}{I({G})},\\
&F_{hybrid} = 2\times \frac{Precision_{E}\times Recall_L}{Precision_E+Recall_{L}}.
\end{split}
\end{equation}

\noindent where $E$, $G$, and $L$ are three sets, containing the detected emerging issues, the key terms in the changelogs, and all app issues (including both emerging and non-emerging issues), respectively. $I(\boldsymbol{\cdot})$ denotes the number of the issues in $\boldsymbol{\cdot}$. We experimentally set the parameters as $\omega=3$, $K=13$, $PMI=5$, $\mu=1.0$, and $m=0.5$. We also initialize $\alpha$ and $\beta$ with 0.1 and 0.01, respectively.

% \begin{table}[]
% 	\centering
% % 	\captionsetup{aboveskip=-0.5pt}
% 	\caption{Changelog of YouTube}
% 	\label{tab:case_c}
% 	\scalebox{0.75}{%
% 		\begin{tabular}{|c|c|m{7.5cm}|}
% 			%       \rowcolor{gray!50}
% 			\thickhline
% 			 Version & Date & Changelog \\
% 			\thickhline
% 			\multirow{5}{*}{11.10} & \multirow{5}{*}{22-Mar-16} & (1) Added \hl{slide over} and \hl{split view} support \\
% 			&  & (2) Moved \hl{home tabs} into \hl{navigation bar} for iPad in \hl{landscape mode} \\
% 			&  & (3) Fixed bug that prevented \hl{URLs} in \hl{video descriptions} from opening\\
% 			\hline
% 			\multirow{5}{*}{11.11} & \multirow{5}{*}{29-Mar-16} & (1) Fixed bug where accessibility \hl{VoiceOver} \hl{looped} over the same elements  \\
% 			&  & (2) Fixed issue where the video couldn't be \hl{exited after completing} \\
% 			&  & (3) Bug fixes and stability improvements \\
% 			\thickhline
% 		\end{tabular}
% 	}
% \end{table}

\subsection{Baseline Approaches}
We compare the effectiveness of our proposed framework with a popular emerging event detection approaches on social networks, OLDA~\cite{DBLP:conf/icdm/AlSumaitBD08} and the state-of-the-art emerging app issue identification approach, IDEA~\cite{gao2018online}.
%  and DIVER~\cite{gao2019diver}
%  and BBTM~\cite{DBLP:conf/aaai/YanGLXC15},

\textbf{On-line Latent Dirichlet Allocation (OLDA)} is an online version of Latent Dirichlet Allocation (LDA)~\cite{DBLP:conf/nips/BleiNJ01} that manually captures the topic patterns and identifies topics of text streams and their changes over time. It generates an evolutionary word distribution matrix for each topic. In this way, it incrementally builds an up-to-date model when new documents appear. The emerging topics in the current app version are determined by comparison with the topic distributions in the previous version.

% \textbf{Bursty Biterm Topic Model (BBTM)} is an extension version of Biterm Topic Model (BTM)~\cite{DBLP:conf/www/YanGLC13} by incorporating the burstiness of biterms as prior knowledge. Different from OLDA~\cite{DBLP:conf/icdm/AlSumaitBD08} which models the words in corpus, BBTM models biterm (\textit{i.e.}, word pairs) and can alleviate the data sparsity problem of short texts. 

\textbf{IDEA} is a state-of-the-art emerging app issue identification approach proposed recently. It improves OLDA by considering the topic distributions in previous versions within a version window during emerging topic detection. The improved method is named as Adaptive OLDA (AOLDA). It also includes an automatic topic interpretation method for labeling each topic with the most representative phrases and sentences.

% combining the previous topics in a version window for emerging topic detection instead of considering the only one previous topics. 

% \textbf{DIVER} is an emerging app issue detection tool applied in industrial scenario. It saves the word collocations and corresponding statistics (\textit{e.g.}, frequency and post time) in a steady interval continuously. For each time slice, DIVER groups the emerging word collocations and presents to developers together with the retrieved relevant reviews. DIVER is claimed to demonstrate better performance than IDEA~\cite{gao2018online} in industry.

\begin{table*}[t]
	\centering
	\caption{Comparison results with baseline approaches. The value under each app name indicates the average number of reviews across the versions, and \textbf{bold} figures highlight better results.}
	\label{tab:merti_idea}
	
	\scalebox{0.9}{%
    	\begin{tabular}{c|c|r|r|r|r|r|r}
    	\hline
    	\hline
    	\multirow{2}{*}{\begin{tabular}[x]{@{}c@{}}\textbf{App Name}\\\textbf{(\#avg. reviews)}\end{tabular}} & \multirow{2}{*}{\textbf{Method}} & \multicolumn{3}{c|}{\textbf{Phrase}} & \multicolumn{3}{c}{\textbf{Sentence}}\\
    	\cline{3-8}
    	& & $Precision_E$ & $Recall_L$ & $F_{hybrid}$ & $Precision_E$ & $Recall_L$ & $F_{hybrid}$ \\
    	\hline
    	\multirow{3}{*}{\begin{tabular}[x]{@{}c@{}}YouTube\\(1,143)\end{tabular}} & OLDA & 0.441 & 0.462 & 0.451	& 0.578	& 0.664	& 0.597 \\
    	& IDEA & 0.592 &	0.472 &	0.523 &	0.628 &	0.666 &	0.636 \\
    	& MERIT & \textbf{0.625} &	\textbf{0.551} &	\textbf{0.586} &	\textbf{0.667} &	\textbf{0.760} &	\textbf{0.710} \\
    	\hline
    	\multirow{3}{*}{\begin{tabular}[x]{@{}c@{}}Clean Master\\(6,332)\end{tabular}} & OLDA & 0.300	& 0.269	 & 0.160 &	0.200	& 0.421 &	0.129 \\
    	& IDEA & \textbf{0.667} &	0.318 &	0.431 &	0.667 &	0.434 &	0.526\\
    	& MERIT & \textbf{0.667} &	\textbf{0.468} &	\textbf{0.550} &	\textbf{0.833} &	\textbf{0.848} &	\textbf{0.841} \\
    	\hline
    	\multirow{3}{*}{\begin{tabular}[x]{@{}c@{}}Viber\\(2,141)\end{tabular}} & OLDA & 0.157 &	0.305 &	0.166 &	0.313 &	0.550 &	0.375\\
    	& IDEA & 0.625 &	0.340 &	0.440 &	0.625 &	0.651 &	0.638 \\
    	& MERIT & \textbf{0.667} &	\textbf{0.706} &	\textbf{0.686} &	\textbf{0.833} &	\textbf{0.809} &	\textbf{0.821} \\
    	\hline
    	\multirow{3}{*}{\begin{tabular}[x]{@{}c@{}}Ebay\\(3,943)\end{tabular}} & OLDA & 0.167 &	0.238	& 0.196 &	0.500 &	0.488 &	0.494\\
    	& IDEA & 0.229 &	0.251 &	0.227 &	0.646 &	0.527 &	0.580 \\
    	& MERIT & \textbf{0.889}	& \textbf{0.508} &	\textbf{0.646} &	\textbf{1.000} &	\textbf{0.749} &	\textbf{0.857} \\
    	\hline
    	\multirow{3}{*}{\begin{tabular}[x]{@{}c@{}}SwiftKey\\(1,313)\end{tabular}} & OLDA & 0.100 & 0.567& 0.148 & 0.367 & 0.617 & 0.458 \\
    	& IDEA & 0.517 &	\textbf{0.653} &	0.523 &	0.583 &	0.700 &	0.587 \\
    	& MERIT & \textbf{0.800}	 & 0.633 &	\textbf{0.707} &	\textbf{0.800} &	\textbf{0.867} &	\textbf{0.832} \\
    	\hline
    	\multirow{3}{*}{\begin{tabular}[x]{@{}c@{}}NOAA Radar\\(523)\end{tabular}} & OLDA & 0.468 & 0.528 & 0.473 & 0.482 & 0.622 & 0.534 \\
    	& IDEA & 0.571 &	0.497 &	0.531 &	0.476 &	0.639 &	0.546 \\
    	& MERIT & \textbf{0.750} &	\textbf{0.654} &	\textbf{0.699} &	\textbf{0.750} &	\textbf{0.840} & \textbf{0.793} \\
    	\hline
    	\hline
    	\end{tabular}
    }
\end{table*}

\section{Experimental Results}\label{sec:experiment}
In this section, we describes results of the evaluation of MERIT through experiments and compare it with the state-of-the-art tool, IDEA~\cite{gao2018online}, and another competing approach, OLDA~\cite{DBLP:conf/icdm/AlSumaitBD08}, to assess its capability in identifying emerging app issues for developers. Our experiments are aimed to answer the following research questions:

%  and DIVER~\cite{gao2019diver}

\begin{enumerate}[label=\bfseries RQ\arabic*:,leftmargin=.5in]
    \item What is the performance of MERIT in detecting emerging app issues? 
    % \item What is the performance of MERIT in industry scenario?
    \item What is the impact of different extensions on the performance of MERIT? The extensions include adopting BTM for topic modeling instead of LDA, considering sentiment for each topic, and the new topic labeling approach.
    % Can online Biterm topic modeling approach outperforms OLDA in topic inference?
    % \item What is the performance of the proposed word-embedding-enhanced topic interpretation method? 
    % \item What is the impact of sentiment on the performance of MERIT?
    
    % \item How efficient is MERIT?
    
    % \item What is the impact of different parameters on the performance of MERIT?
    
\end{enumerate}

\subsection{RQ1: What is the performance of MERIT in detecting emerging app issues?} 
This research question relates to the capability of MERIT in identifying accurate and complete emerging app issues in comparison with IDEA~\cite{gao2018online} and OLDA~\cite{DBLP:conf/icdm/AlSumaitBD08}. Having too many false positives would end up being counterproductive, whereas having too many false negatives would mean that the proposed framework is not able to alert emerging issues in many cases where those are important. Table~\ref{tab:merti_idea} displays the comparison results.

As seen in Table~\ref{tab:merti_idea}, the proposed MERIT approach outperforms the baseline approaches on all the metrics. We discuss the performance of MERIT from two aspects as below.

\textbf{\yun{Result} 1: Interpreting Topics with Phrases v.s. Sentences.} As mentioned in Section~\ref{sssec:topic_inter}, there are two ways to represent an app issue: by phrases and by sentences. For example, the ``Label'' and ``Sentences'' in the top boxes of Figure~\ref{fig:river} are the phrase and sentence representations respectively. As shown in Table~\ref{tab:merti_idea}, considering all the three methods, issues in sentences present better performance than those in phrases with 9.5\%, 29.1\%, and 15.6\% increase in $Precision_E$, $Recall_L$, and $F_{hybrid}$ on average respectively. This result may be attributed to the fact that sentences can convey more details than phrases and thereby cover more key terms mentioned in changelogs, which is also in line with findings of our previous study~\cite{gao2018online}. Specifically, the sentences identified by MERIT can enhance the performance of phrases by 8.1\%, 22.6\%, and 16.3\% wrt. three metrics, respectively. We then use Wilcoxon signed-rank test~\cite{wilcoxon1992individual} for statistical significance test, and Cliff's Delta (or $d$) to measure the effect size~\cite{ahmed2006effect}. The significance test result ($p-value<0.05$) and large effect size ($d=2.76$) on the difference in the mean of the $F_{hybrid}$ scores of phrase-level issues and sentence-level issues confirm the better performance of sentence representations over phrase representations.

\textbf{Result 2: MERIT v.s. Baselines.} Comparing MERIT with baseline approaches, we find that MERIT can outperform both baselines in all the three metrics with respect to sentence-level issues. For phrase-level issues, although MERIT shows a slightly lower $Recall_L$ than IDEA for the SwiftKey app, it exhibits better performance in both $Precision_E$ and $F_{hybrid}$. On average, MERIT can achieve precision, recall, and f-score of 81.4\%, 81.2\%, and 80.9\% respectively, and outperform OLDA by 37.8\% and IDEA by 22.3\% for $F_{hybrid}$, which indicates that MERIT can better balance the precision and recall in emerging issue detection. Besides, the significant statistical test results ($p-value<0.01$) and large effect sizes ($d>2$) on the $F_{hybrid}$ scores for both phrase and sentence -level issues of MERIT and IDEA/OLDA confirm the superiority of MERIT over IDEA/OLDA.

\subsection{RQ2: What is the impact of different extensions on the performance of MERIT?}
MERIT extends IDEA by (1) adopting BTM for topic modeling instead of LDA, (2) jointly modeling sentiment and topics, and (3) employ the proposed word-embedding-based topic labeling (NTL) approach. We perform ablation
% contrastive 
experiments by considering each of the 3 extensions 
one-at-a-time, 
% and create three other approaches built upon IDEA, 
which we refer to as
% within which only a single component is augmented to the 
``+BTM'', ``+Sentiment'', and ``+NTL'' respectively. Table~\ref{tab:word2vec_res} shows the results of comparing each of these 3 approaches with the baselines.

Unsurprisingly, the combination of all extensions gives the greatest improvements in terms of $F_{hybrid}$, and all the components are beneficial on their own. Similar to the answer to RQ1, we also observe that sentence-level issues generally present better performance than the phrase-level issues. 
% The inverse pattern only happens for $Precision_E$ of the NOAA Radar app, where phrases can more precisely represent the emerging issues than sentences. We guess that such a case may be because the small volume of reviews for the app (523 reviews per version) \yun{to add@chuan}.
% weakens the effect of the augmented components. 

Specifically, with respect to each extension 
considered independently using BTM instead of LDA for topic modeling can enhance the average performance by 8.8\% and 16.9\% for the phrase-level and sentence-level $F_{hybrid}$ scores respectively. \yun{In terms of $Precision_{E}$ and $Recall_{L}$, with BTM involved, the performance increases by 11.5\% and 19.0\%, respectively.} When jointly modeling topics with \yun{the} sentiment, the $F_{hybrid}$ scores are increased by 5.8\% and 19.4\% in terms of phrase and sentence representations, respectively. \yun{On average, both precision and recall show an increasing trend, +9.6\% and +13.4\%, respectively. The results indicate that by the considerations of sentiments, overall results including both precision and recall have been improved. But for some apps, such as YouTube and Clean Master, although the recall is increased (+20.2\% and +31.4\% respectively), the precision is slightly dropped (-7.8\% and -4.2\% respectively). This may be because with the sentiment involved, the topics predicted as negative sentiment tend to be identified as emerging issues, which is helpful for enhancing the recall. But the negative topics might not always be emerging, such as some constantly recurring topics (e.g., the ``\textit{screen}'' topic for YouTube and the ``\textit{battery}'' topic for Clean Master), so the precision is slightly weakened. Besides}, involving word embeddings during topic interpretation gives us a 7.3\% increase for phrase-level issues and a 5.9\% increase for sentence-level issues with respect to $F_{hybrid}$. \yun{We also observe that although the YouTube app (with 1,143 reviews per version) shows a slightly decrease (-2.4\%) on the $F_{hybrid}$ score, all the other apps, especially NOAA Radar which has only 523 reviews per version, enjoy an increase. Thus, the experiment results demonstrate that the novel topic labeling method can work well even for apps with few reviews.} Moreover, the gain from different extensions is not fully cumulative since the information delivered by these components overlaps. For instance, both the topic modeling and topic labeling steps help capture the semantics of the words in app reviews to generate accurate emerging issues.

\begin{table*}[h]
	\centering
	\caption{Ablation experiments with different extensions turned on/off. The value under each app name indicates the average number of reviews across the versions, and \textbf{bold} figures highlight better results. The methods ``+BTM'', ``+Sentiment'', ``+NTL'' respectively represent the extensions upon IDEA that we propose in this work, including using BTM for topic modeling instead of LDA, combining topics with sentiment, and enhancing the topic labeling step with word embeddings.}
	\label{tab:word2vec_res}
	
	\scalebox{0.9}{%
    	\begin{tabular}{c|c|r|r|r|r|r|r}
    	\hline
    	\hline
    	\multirow{2}{*}{\begin{tabular}[x]{@{}c@{}}\textbf{App Name}\\\textbf{(\#avg. reviews)}\end{tabular}} & \multirow{2}{*}{\textbf{Method}} & \multicolumn{3}{c|}{\textbf{Phrase}} & \multicolumn{3}{c}{\textbf{Sentence}}\\
    	\cline{3-8}
    	& & $Precision_E$ & $Recall_L$ & $F_{hybrid}$ & $Precision_E$ & $Recall_L$ & $F_{hybrid}$ \\
    	\hline
    	\multirow{5}{*}{\begin{tabular}[x]{@{}c@{}}YouTube\\(1,143)\end{tabular}} &  IDEA & 0.592 &	0.472 &	0.523 &	0.628 &	0.666 &	0.636 \\
    	\cdashline{3-8}
    	& +BTM & 0.525 &	0.576 &	0.416 &	0.592 &	0.843 &	0.696 \\
    	& +Sentiment & 0.483 &	\textbf{0.592} &	0.532 &	0.550 &	\textbf{0.868} &	0.673 \\
    	& +NTL & 0.544 &	0.477 &	0.523 &	0.582 &	0.773 &	0.612 \\
    	\cdashline{3-8}
    	& MERIT & \textbf{0.625} &	0.551 &	\textbf{0.586} &	\textbf{0.667} &	0.760 &	\textbf{0.710} \\
    	\hline
    	\multirow{5}{*}{\begin{tabular}[x]{@{}c@{}}Clean Master\\(6,332)\end{tabular}} & IDEA & 0.667 &	0.318 &	0.431 &	0.667 &	0.434 &	0.526 \\
    	\cdashline{3-8}
    	& +BTM & 0.444 &	0.420 &	0.432 &	0.778 &	0.761 &	0.769\\
    	& +Sentiment & 0.417 &	0.335 &	0.371 &	0.625 &	0.748 &	0.681 \\
    	& +NTL & \textbf{0.833} &	0.299 &	0.440 &	0.747 &	0.500 &	0.599 \\
    	\cdashline{3-8}
    	& MERIT & 0.667 &	\textbf{0.468} &	\textbf{0.550} &	\textbf{0.833} &	\textbf{0.848} &	\textbf{0.841} \\
    	\hline
    	\multirow{5}{*}{\begin{tabular}[x]{@{}c@{}}Viber\\(2,141)\end{tabular}} & IDEA & 0.625 &	0.340 &	0.440 &	0.625 &	0.651 &	0.638 \\
    	\cdashline{3-8}
    	& +BTM & 0.625 &	0.692 &	0.657 &	0.750 &	\textbf{0.809} &	0.778\\
    	& +Sentiment & \textbf{0.778} &	0.395 &	0.524 &	0.778 &	0.566 &	0.655\\
    	& +NTL & 0.667 &	0.366 &	0.473 &	0.664 &	0.778 &	0.716 \\
    	\cdashline{3-8}
    	& MERIT & 0.667 &	\textbf{0.706} &	\textbf{0.686} &	\textbf{0.833} &	\textbf{0.809} &	\textbf{0.821} \\
    	\hline
    	\multirow{5}{*}{\begin{tabular}[x]{@{}c@{}}Ebay\\(3,943)\end{tabular}} & IDEA & 0.229 &	0.251 &	0.227 &	0.646 &	0.527 &	0.580\\
    	\cdashline{3-8}
    	& +BTM & 0.667 &	0.402 &	0.502 &	0.833 &	0.640 &	0.780 \\
    	& +Sentiment & 0.361 &	0.310 &	0.333 &	0.833 &	0.516 &	0.637 \\
    	& +NTL & 0.542 &	0.418 &	0.472 &	0.676 &	0.667 &	0.671 \\
    	\cdashline{3-8}
    	& MERIT & \textbf{0.889}	& \textbf{0.508} &	\textbf{0.646} &	\textbf{1.000} &	\textbf{0.749} &	\textbf{0.857} \\
    	\hline
    	\multirow{5}{*}{\begin{tabular}[x]{@{}c@{}}SwiftKey\\(1,313)\end{tabular}} & IDEA & 0.517 &	0.653 &	0.523 &	0.583 &	0.700 &	0.587 \\
    	\cdashline{3-8}
    	& +BTM & 0.500 &	\textbf{0.767} &	0.605 &	0.750 &	\textbf{0.900} &	0.818 \\
    	& +Sentiment & 0.500 &	0.667 &	0.571 &	0.750 &	0.867 &	0.704 \\
    	& +NTL & 0.500 &	0.733 &	0.595 &	0.500 &	0.767 &	0.605 \\
    	\cdashline{3-8}
    	& MERIT & \textbf{0.800}	 & 0.633 &	\textbf{0.707} &	\textbf{0.800} &	0.867 &	\textbf{0.832} \\
    	\hline
    	\multirow{5}{*}{\begin{tabular}[x]{@{}c@{}}NOAA Radar\\(523)\end{tabular}} & IDEA & 0.571 &	0.497 &	0.531 &	0.476 &	0.639 &	0.546 \\
    	\cdashline{3-8}
    	& +BTM & 0.611 &	0.575 &	0.592 &	0.611 &	0.773 &	0.683 \\
    	& +Sentiment & \textbf{0.796} &	0.612 &	0.692 &	0.667 &0.829 & 0.739 \\
    	& +NTL & 0.667 &	0.566 &	0.612 &	0.619 &	0.710 &	0.662 \\
    	\cdashline{3-8}
    	& MERIT & 0.750 &	\textbf{0.654} &	\textbf{0.699} &	\textbf{0.750} &	\textbf{0.840} & \textbf{0.793} \\
    	\hline
    	\hline
    	\end{tabular}
    }
\end{table*}

\section{Discussions}\label{sec:limitation}
In this section, we discuss the advantages of MERIT, its limitations, \yun{time cost,} impacts of different parameters, and the threats.

\subsection{Why does Our Model Work?}
We have identified three advantages of MERIT that may explain its effectiveness in detecting emerging app issues.

\textbf{Observation 1: MERIT can better model the topics of short texts.}
In this work, we propose to use the biterm topic model (BTM) for short text mining instead of LDA. Since LDA learns review-level word co-occurrence patterns to reveal topics, it suffers from the severe data sparsity in short review texts. Instead, BTM learns the topics from word co-occurrence patterns directly and thus alleviate the data sparsity problem. Table~\ref{tab:topic_words_com} shows the top eight terms of three example topics obtained from LDA and BTM. We discover that BTM can generate more semantically-coherent terms for each topic. For example, the terms ``\textit{also}'', ``\textit{there$\_$be}'', and ``\textit{great$\_$app}'' terms are not related to Topic 2 ``\textit{play button}''. The semantic inconsistency of the top terms in one emerging issue would confuse developers or influence the performance of subsequent automatic topic interpretation step. By using BTM instead of LDA for review modeling, the semantics of top terms belonging to one topic can be more coherent.

\begin{table}[h]
	\center
	\caption{Comparison on the topics generated by LDA and BTM for the YouTube iOS app. The topics are related to ``\textit{battery drainage}'', ``\textit{play button}'', and ``\textit{video recommendation}'' respectively, each with top eight terms presented. \yun{Fonts with wavy underlines} highlight the terms that are not semantically related to the issue topic.}
	\label{tab:topic_words_com}
	\scalebox{0.9}{
	\begin{threeparttable}
	\begin{tabular}{l| c|c|c}
		\hline
		 \multirow{2}{*}{\textbf{Method}} & \textbf{Topic 1} & \textbf{Topic 2} & \textbf{Topic 3}  \\
		 & \textit{Battery drainage} & \textit{Play button} & \textit{Video Recommendation} \\
		 \hline
		 \multirow{8}{*}{\textbf{LDA}} & $\text{\textless digit\textgreater}$& video  & video \\
		 & \uwave{quality}  &problem  & watch \\
		 & battery & \uwave{also}& see \\
		 & \uwave{io} & make & channel \\
		 & iphone & button  & recommend \\
		 & use & \uwave{$\text{there}\_\text{be}$} & find \\
		 & 6s & go & \uwave{anymore} \\
		 & video &\uwave{$\text{great}\_\text{app}$} & \uwave{$\text{i}\_\text{want}$}\\
		 \hline
		 \multirow{8}{*}{\textbf{BTM}} & $\text{\textless digit\textgreater}$& video& video\\
		 & battery & play& home\\
		 & video & button & watch\\
		 & drain & screen & screen\\
		 & use &auto & page \\
		 & cause & arrow& thumbnail\\
		 & 6s & watch& see\\
		 & watch & pause & recommend\\
		 \hline
		 
		 \hline
	\end{tabular}
% 	\begin{tablenotes}
%     \item[1] Measured in percentages.
%   \end{tablenotes}
	\end{threeparttable}
}
\end{table}

\textbf{Observation 2: MERIT can focus on negative topics.} As shown in Figure~\ref{fig:btm_sample}, the topics extracted from reviews are usually mingled with various polarities. Even for the same topics, users may express totally different opinions. Motivated by the intuition that developers are more concerned about the negative app aspects~\cite{chen2014ar,guzman2014users}, MERIT focuses on the topics inferred as negative instead of incorporating topics in all sentiment polarities. Thus, MERIT can expose the topics likely corresponding to app issues.

\textbf{Observation 3: MERIT can interpret topics with more representative and coherent labels.} For accurate topic labeling, we combine word embeddings to prioritize semantically-representative phrase/sentence candidates. Table~\ref{tab:tl_result} shows the ranked phrases with and without word embeddings involved, respectively. We can discover that the proposed word-embedding-enhanced topic labeling (NTL) approach can better interpret the topic meanings in terms of coherence and semantic accuracy. For example, the original topic labeling approach selects ``\textit{playback error galore}'' and ``\textit{playback error}'' as the most representative phrases of Topic 1 and Topic 2 respectively, which are intuitively different from the general meaning of the topics (\textit{i.e.}, split view and subscription box respectively) and can cause confusion to developers in understanding the detected emerging issues. Instead, the top three phrases of these two topics obtained by MERIT are semantically coherent, all about split view or subscription.

\begin{table}[h]
	\center
	\caption{Examples of ranked phrases with the original topic labeling approach (denoted as MERIT+TL) and the new word-embedding-based approach (denoted as MERIT+NTL). \yun{Fonts with wavy underlines} highlight the phrase labels that are not semantically related to the corresponding topic. We also present the ground truth corresponding to each topic in boldface.}
	\label{tab:tl_result}
	\scalebox{0.76}{
	\begin{tabular}{>{\centering\arraybackslash}m{0.09\textwidth}| >{\centering\arraybackslash}m{0.17\textwidth}|>{\centering\arraybackslash}m{0.14\textwidth}|c}
		\hline
		 \multirow{3}{*}{} & \textbf{Topic 1} & \textbf{Topic 2} & \textbf{Topic 3} \\
		 & \begin{tabular}[x]{@{}c@{}}\textit{Third party utility}\\\textit{for split view}\end{tabular} &\textit{Subscription box} & \textit{Comment section}\\
		 \hline
		 \multirow{8}{*}{\textbf{Top Terms}} & picture & watch & comment\\
		 & support & video & video \\
		 & feature &say & change \\
		 & add & subscription & better \\
		 & video & show& description\\
		 & $\text{slide}\_\text{over}$ & even & $\text{comment}\_\text{section}$\\
		 & ipad & mark & back \\
		 & $\text{split}\_\text{screen}$ & see & move \\
		 \hline
		 \multirow{3}{*}{\begin{tabular}[x]{@{}c@{}}\textbf{MERIT}\\\textbf{+TL}\end{tabular}}  & \uwave{playback error galore} & \uwave{playback error}& comment section\\
		  & split screen & sub box& \uwave{character limit}\\
		   & browser base & \uwave{low quality} & \uwave{push notification}\\
		 \hline
		 \multirow{3}{*}{\begin{tabular}[x]{@{}c@{}}\textbf{MERIT}\\\textbf{+NTL}\end{tabular}}& split view & subscription fee& channel name\\
		 & split screen & sub box& comment section \\
		 & third party & subscription box& main page\\
		 \hline
		 \begin{tabular}[x]{@{}c@{}}\textbf{Ground}\\\textbf{Truth}\end{tabular} &  \begin{tabular}[x]{@{}c@{}}Added slide over\\ and \textbf{split view}\\ support.\end{tabular} & \begin{tabular}[x]{@{}c@{}}Easier access to \\your full \\\textbf{subscription list}.\end{tabular}& \begin{tabular}[x]{@{}c@{}}Click on \\ timestamp links \\ in \textbf{comments} \\advances the \\ \textbf{video} to \\correct position.\end{tabular}\\
		 \hline

	\end{tabular}
}
\end{table}

% \textbf{Observation 4: MERIT can automatically track the sentiment changes of the topics.}

\subsection{Why does Our Model Fail?}
We have also summarized two main scenarios that may lead to inaccurate emerging issue prediction.

\textbf{Observation 1: MERIT may miss the emerging app issues only mentioned in few user reviews.} For the emerging issues only expressed in few (\textit{e.g.}, three or four) reviews, they are difficult to be exposed through topic modeling approaches~\cite{DBLP:conf/aaai/YanGLXC15,DBLP:journals/www/HuangPWCGZ17}. For example, one major modification claimed by the version 5.9.3 of the Clean Master app, \textit{i.e.}, ``\textit{Added Cloud Recycle Bin - Recover misdeleted photos from the cloud up to 30 days after deleting}'', MERIT misses capturing any emerging issue related to ``\textit{recycle bin}''. After inspecting the collected corpus, we find that only three reviews received in the previous version are describing the recycle bin, which possibly leading to the omission. We discover similar failure scenarios for other apps. For instance, for the NOAA Radar iOS app, it made a major change about its widget in its version 2.0 that fixes an issue, \textit{i.e.}, ``\textit{Tap on Today tab, scroll to the bottom and tap Edit}'' as written in the changelog. MERIT fails to identify the issue since it is only discussed by three pieces of reviews in the corpus of the previous version. 

\textbf{Observation 2: Official changelogs may not cover all the app issues of the previous version that are fixed in the current version.} Although app markets such as the App Store encourages app developers to write what is actually happening to the apps in the changelog~\cite{appstorereleasenote}, app developers tend to write sketchy and vague bullet points for the changes, such as ``\textit{Bug fixes}'' and ``\textit{We're always trying to improve your experience}''. Although we already filter such changelogs out during validation, the release notes may not cover all the major changes made to current versions and could lead to false negatives. 

First, the app issues may not be fixed instantly in the next updated version. For example, MERIT detects an emerging issue associated with video orientation modes, described as ``\textit{portrait mode}'' and ``\textit{full screen mode}'', for version 11.39 of the YouTube iOS app. One user complained that ``\textit{Sometimes when I'm on full screen mode, I click the minimise screen button and it doesn't work. I try to flip my phone and it doesn't minimise.}'', and gave a two-star rating. The issue also aroused heated discussion on the YouTube online forum~\cite{youtubeforum}. We discover that the issue was fixed in a later version 12.05 instead of the immediate next version, as indicated in the changelog ``\textit{Fixed delay when pressing the full screen and minimize buttons in the player。}''. Such postponement of issue fixing is reasonable since not all bugs in apps would be addressed right away~\cite{DBLP:conf/icse/GuoZNM10,DBLP:conf/icsm/ThungLJLRD12}. Second, changelogs may describe modifications in general terms. For example, MERIT alerts an emerging issue related to ``\textit{Samsung keyboard}'' and ``\textit{force close}'' for version 5.0.4.93 of the SwiftKey app. Although the issue is greatly relevant to the corresponding changelog ``\textit{Fixed issues causing repeated crashes on some devices when loading the keyboard}'', the evaluation of MERIT regards ``\textit{device}'' and ``\textit{Samsung}'' as mismatched. Finally, changelogs may not cover all the major app changes. For instance, we find that the voice dictation issue identified by MERIT for version 5.1.0.60 of the SwiftKey app is a change (\textit{i.e.}, forcing install additional app for voice dictation) made in the app version but not described in the changelog. For example, one user commented that ``\textit{After update, force install additional app `google voice search' for voice dictation, which was not previously required. .... I wish I knew so I would not update the app.}''.

% Thereby force unnecessarily to install additional app despite the opposition of the users who only take up space and without which work well in the past. 

% due to their low similarity (0.43)

% not reflect all the major issues concerned by users. For example, MERIT alerts an emerging issue related to ``\textit{Samsung keyboard}'' and ``\textit{enter key}'' for version 5.0.4.93 of the SwiftKey app, \textit{e.g.}, one four-star-rating review stated that ``\textit{Please change keypress sound like on Samsung. Thanks.}''. Although such issue was not mentioned in the changelog, we find that many users complained about it on the online forum~\cite{swiftkeyforum}.

% dislike button'' for the version 11.28 of the YouTube iOS app. One user complained that ``\textit{The comment section was once a fine thing where I could just read comment and simply press like and read reply! Now it is just a clutter mess with the necessary dislike button and it just looks ugly!}'', and gave one-star rating. The complaint was caused by the added dislike button to the comment section for that version, and aroused heated discussion on the official YouTube online forum~\cite{youtubeforum}. 

\subsection{\yun{Efficiency of MERIT}}
\begin{figure}[t]
% \vspace{1mm}
    \centering
    \resizebox{0.95\linewidth}{!}{
    \includegraphics{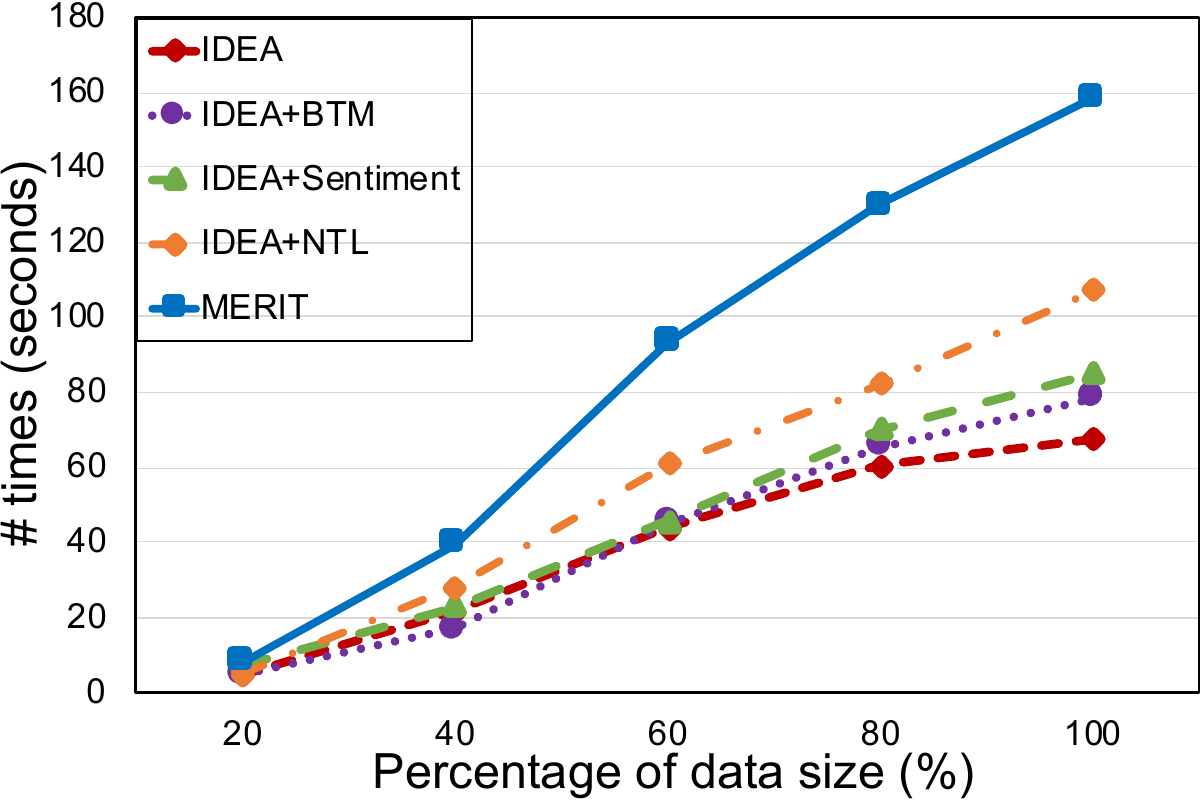}
    }
    \caption{Efficiency of MERIT and the comparison models on different data sizes of 5,000 reviews.}
    \label{fig:run_time}
\end{figure}

\yun{We evaluate whether MERIT can output emerging app issues within reasonable time, by comparing the execution time of MERIT on the subject apps with IDEA, and also with different extensions of IDEA (\textit{i.e.}, ``+BTM'', ``+Sentiment'', and ``+NTL'').} In this experiment, we randomly select subsets of the 5,000 reviews from the YouTube dataset (of different sizes) and run all the models. We run our experiments on a PC with Intel(R) Xeon E5-2620v2 CPU (2.10 GHz, 6 cores) and 16GB RAM. Figure~\ref{fig:run_time} displays the comparison results of time consumed on different dataset sizes. As can be seen from Figure~\ref{fig:run_time}, all the models spend more time as the amount of data increases. We also find that the ``IDEA+BTM'', ``IDEA+Sentiment'' and "IDEA+NTL" models cost 16.0\%, 25.8\%, and 58.4\% more time than the IDEA model when handling 5,000 reviews, respectively. Undoubtedly, MERIT incurs the highest time cost among all the models due to its higher complexity, which can cost 1.3 times more time than IDEA when processing the 5,000 reviews. In spite of the higher time cost, MERIT can deal with 1,000 reviews within eight seconds and 5,000 reviews within three minutes, which we believe to be still acceptable. Therefore, our experiments demonstrate that MERIT can detect emerging app issues more accurately while preserving reasonable time costs.

\subsection{Parameter Analysis}
We also quantitatively compare the performance of MERIT in different parameter settings. We analyze three parameters, that is, the number of topics $K$, the window size $\omega$, the penalty factor $\mu$ (in Equ.~\ref{equ:mu}), and balance parameter $m$ (in Equ.~\ref{equ:labeling}). We vary the values of these four parameters and evaluate their impact on the performance of MERIT. The results are shown in Figure~\ref{fig:parameter}.

\subsubsection{The Number of Topics.} 
\begin{figure*}[ht]
    \centering
    \begin{tabular}{@{\hskip0.5pt}c@{\hskip0.5pt} @{\hskip0.5pt}c@{\hskip0.5pt} @{\hskip0.5pt}c@{\hskip0.5pt} @{\hskip0.5pt}c@{\hskip0.5pt} @{\hskip0.5pt}c@{\hskip0.5pt} @{\hskip0.5pt}c@{\hskip0.5pt}}
        \includegraphics[width=0.16\textwidth]{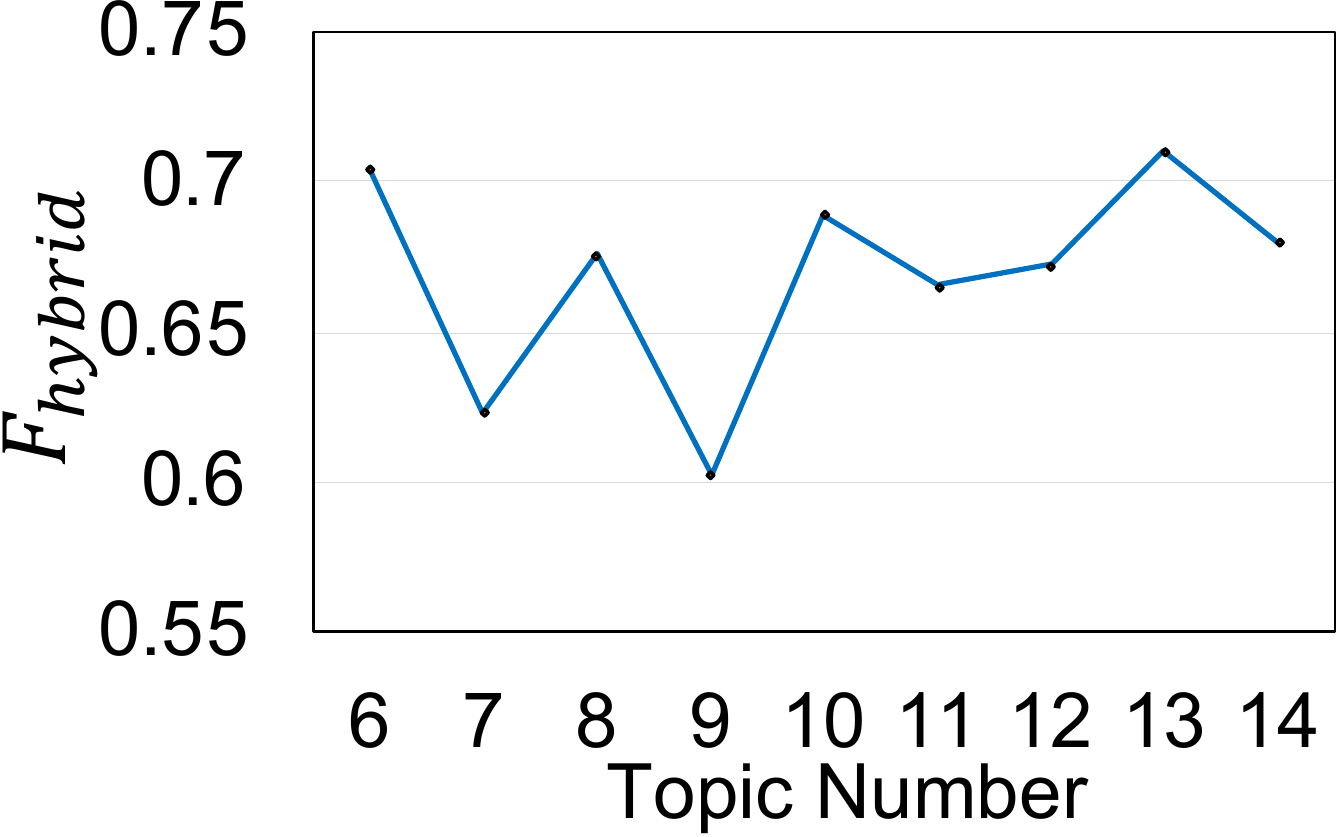} &
        \includegraphics[width=0.16\textwidth]{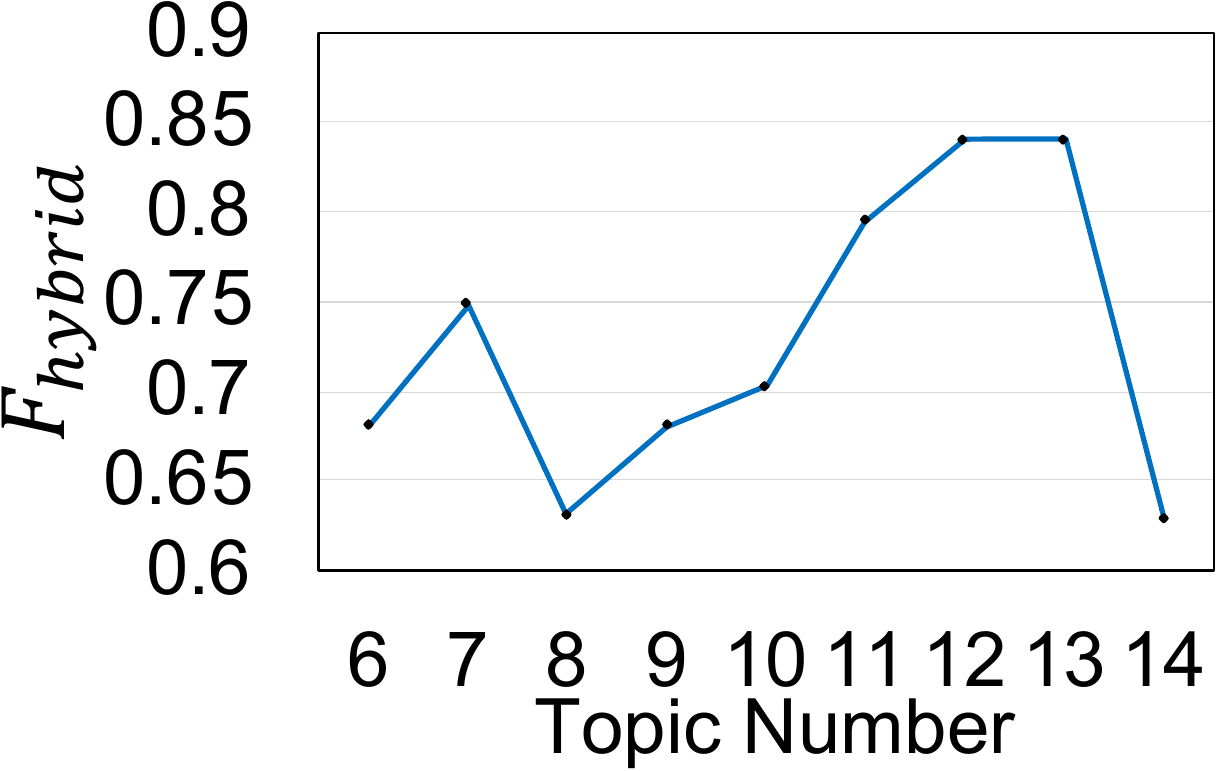} &
        \includegraphics[width=0.16\textwidth]{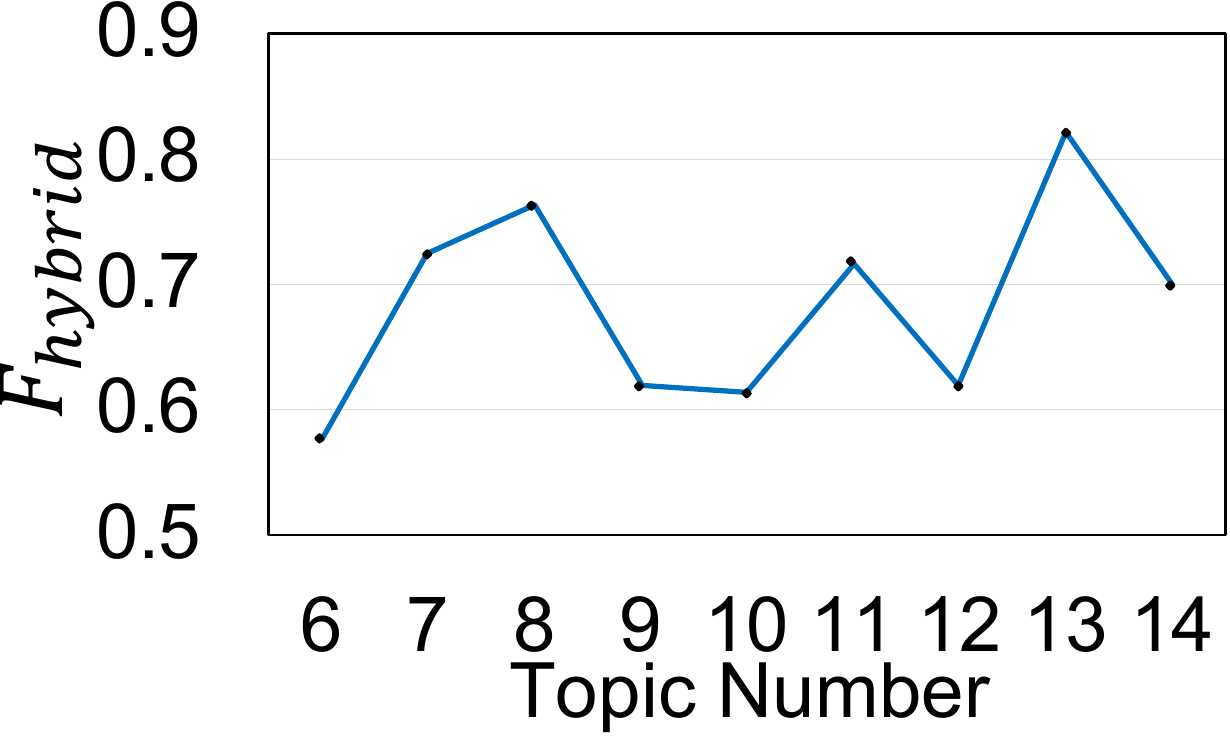} &
        \includegraphics[width=0.16\textwidth]{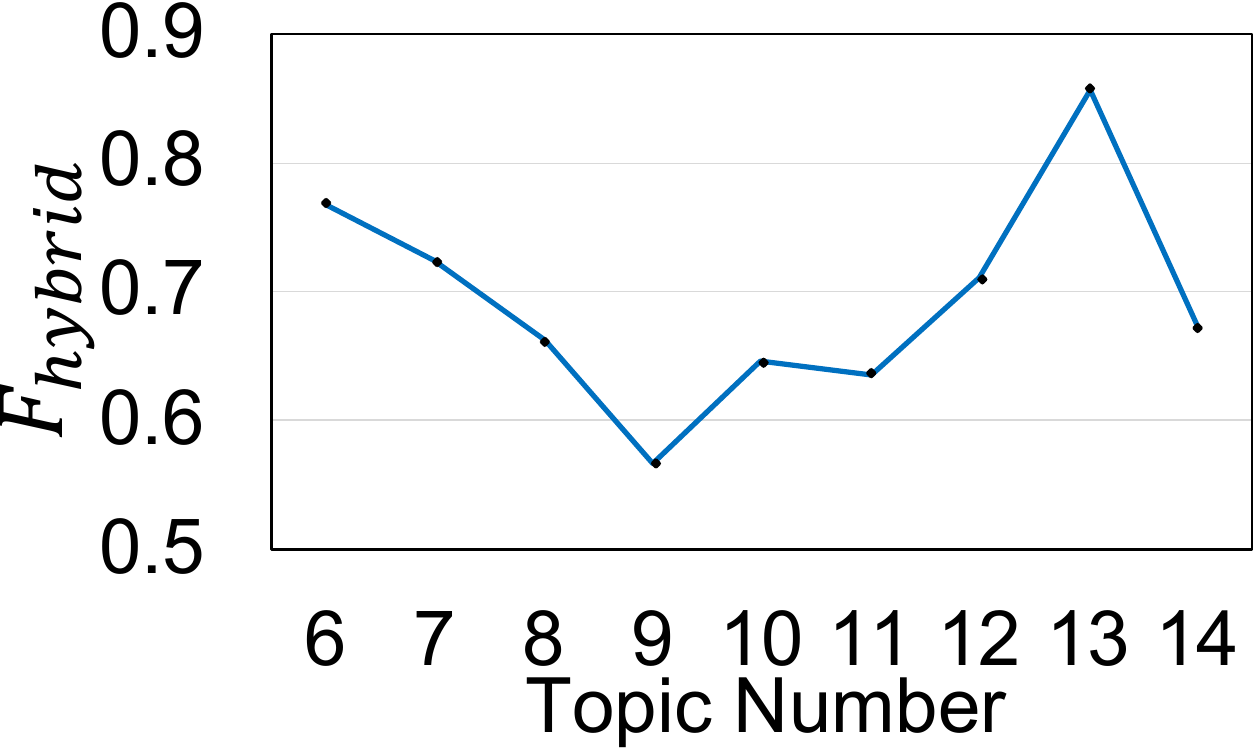} &
        \includegraphics[width=0.16\textwidth]{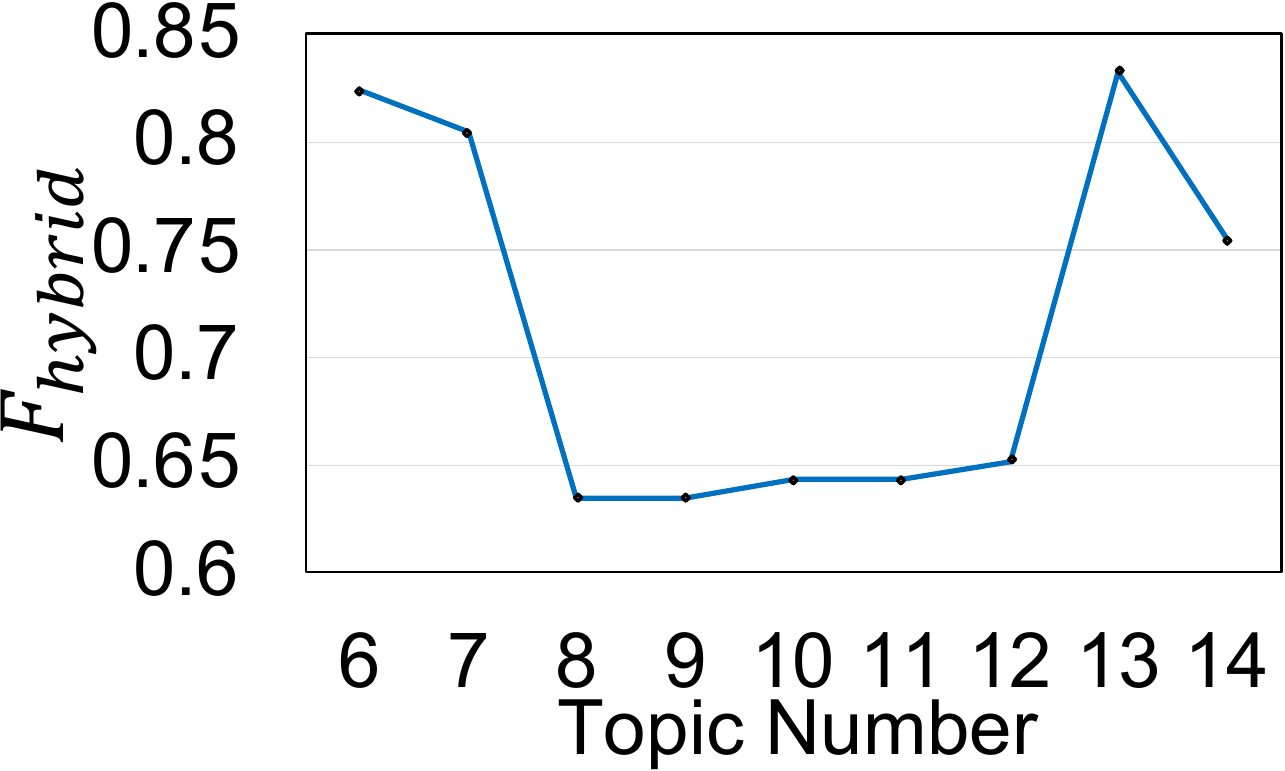} &
        \includegraphics[width=0.16\textwidth]{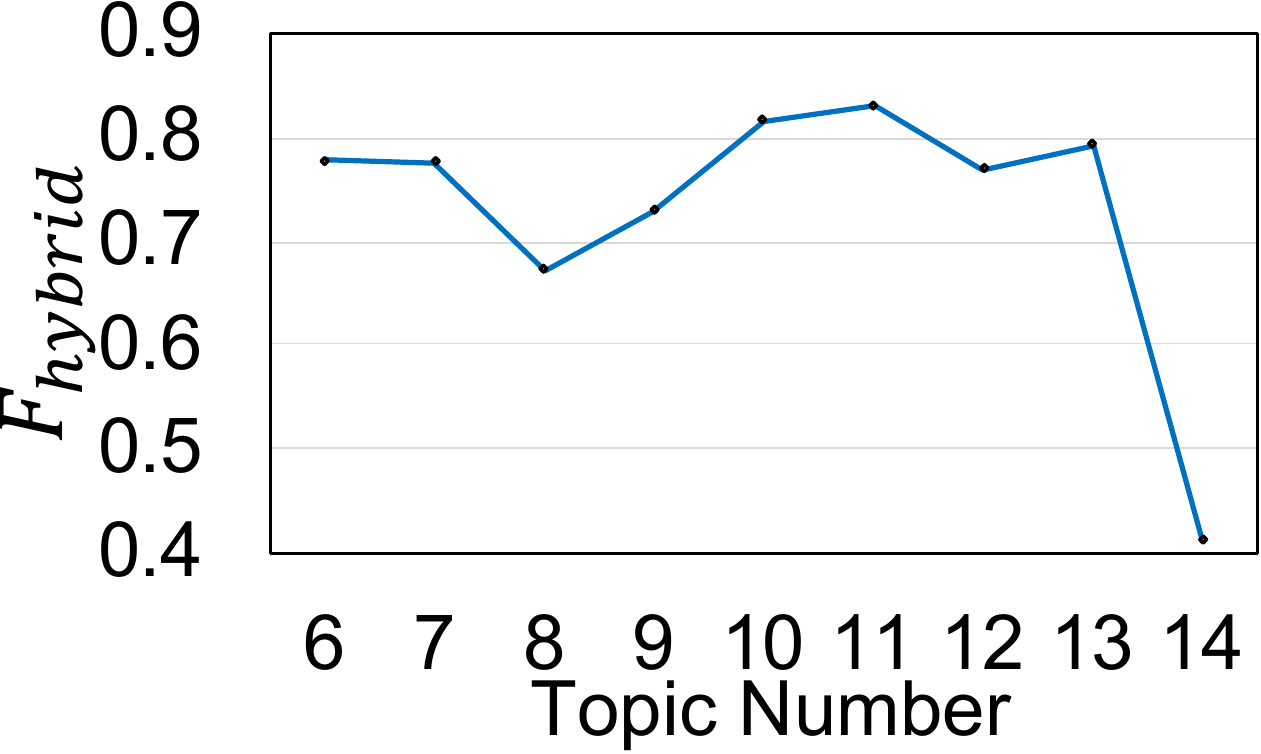} \\
    {\footnotesize (a) YouTube} & {\footnotesize (b) Clean Master} & {\footnotesize (c) Viber} & {\footnotesize (d) Ebay} & {\footnotesize (e) SwiftKey} & {\footnotesize (f) NOAA Radar} \\
    \multicolumn{6}{c}{{\small (1) Topic Number.}} \\
    \includegraphics[width=0.17\textwidth]{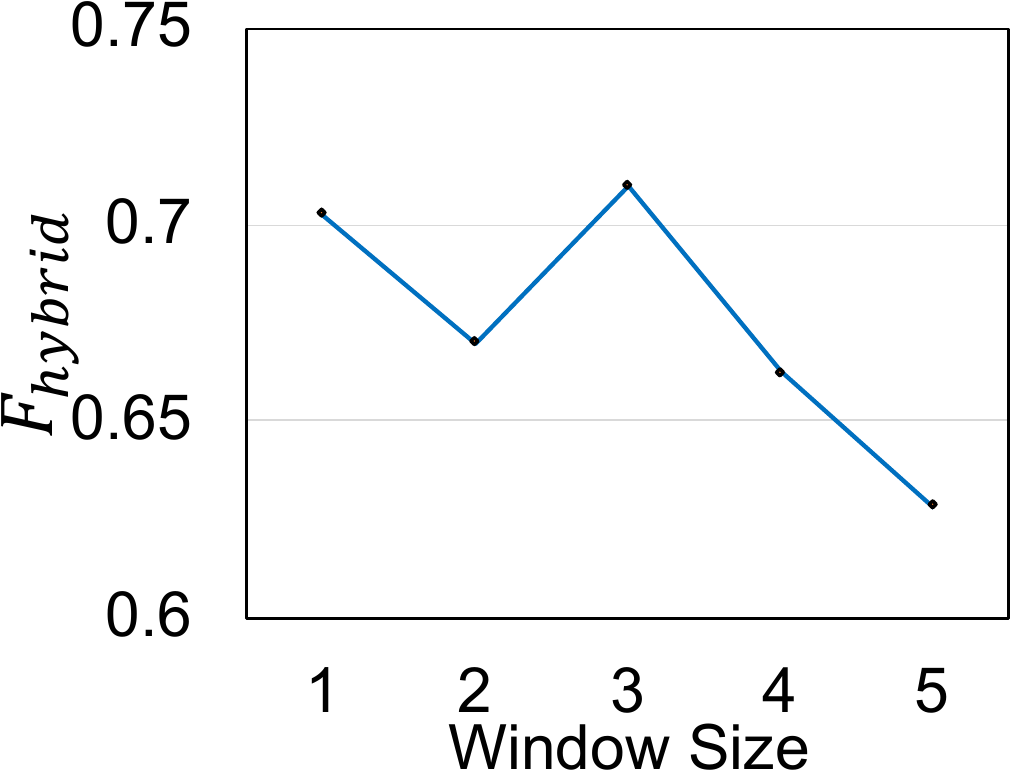} &
    \includegraphics[width=0.17\textwidth]{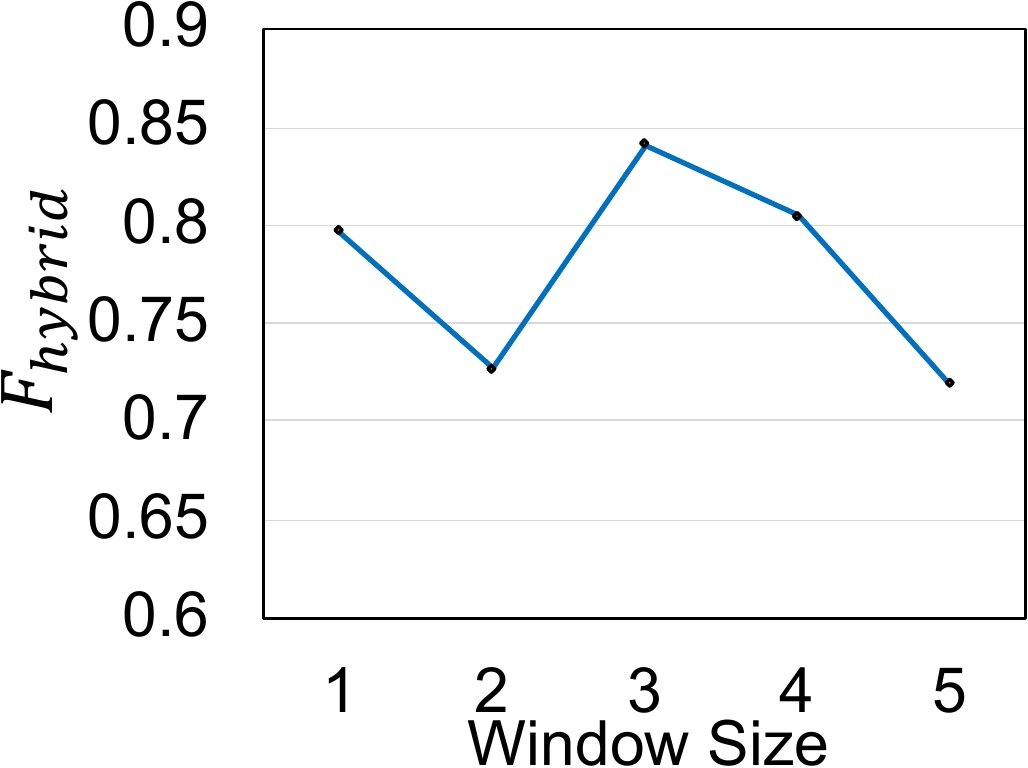} &
    \includegraphics[width=0.16\textwidth]{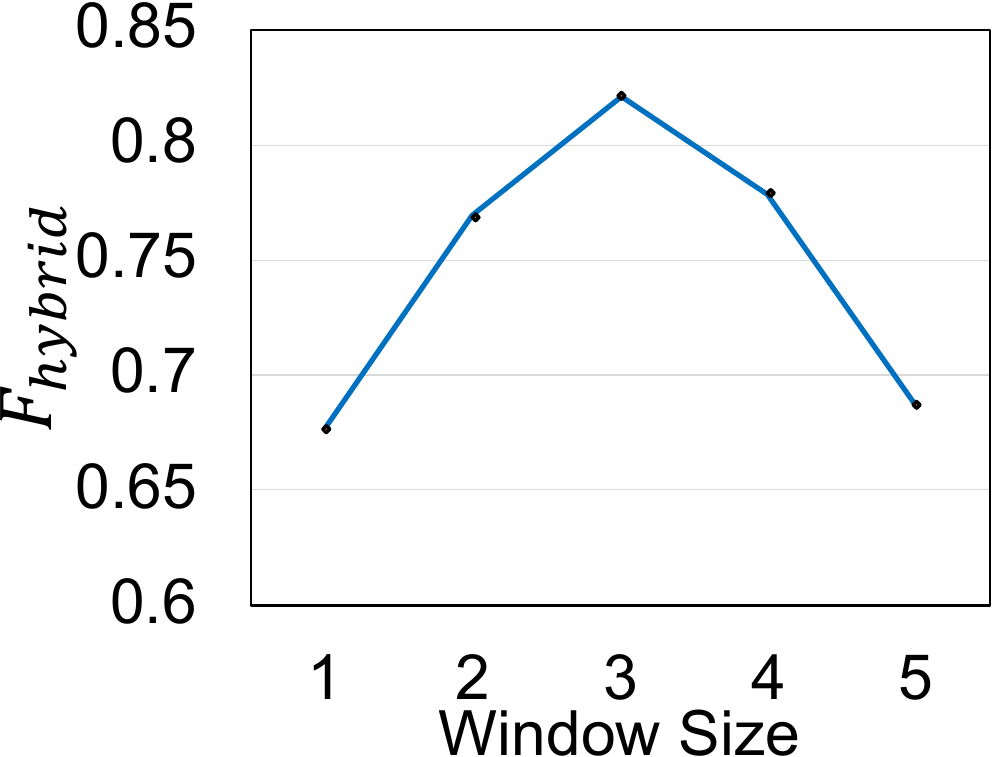} &
    \includegraphics[width=0.16\textwidth]{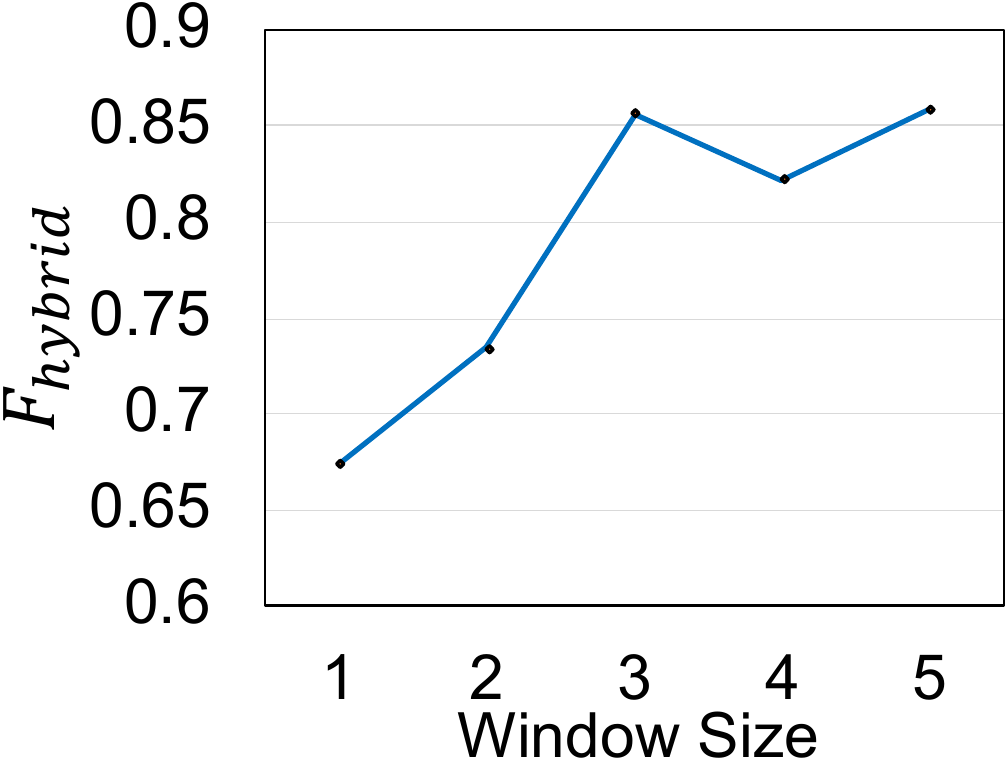} &
    \includegraphics[width=0.15\textwidth]{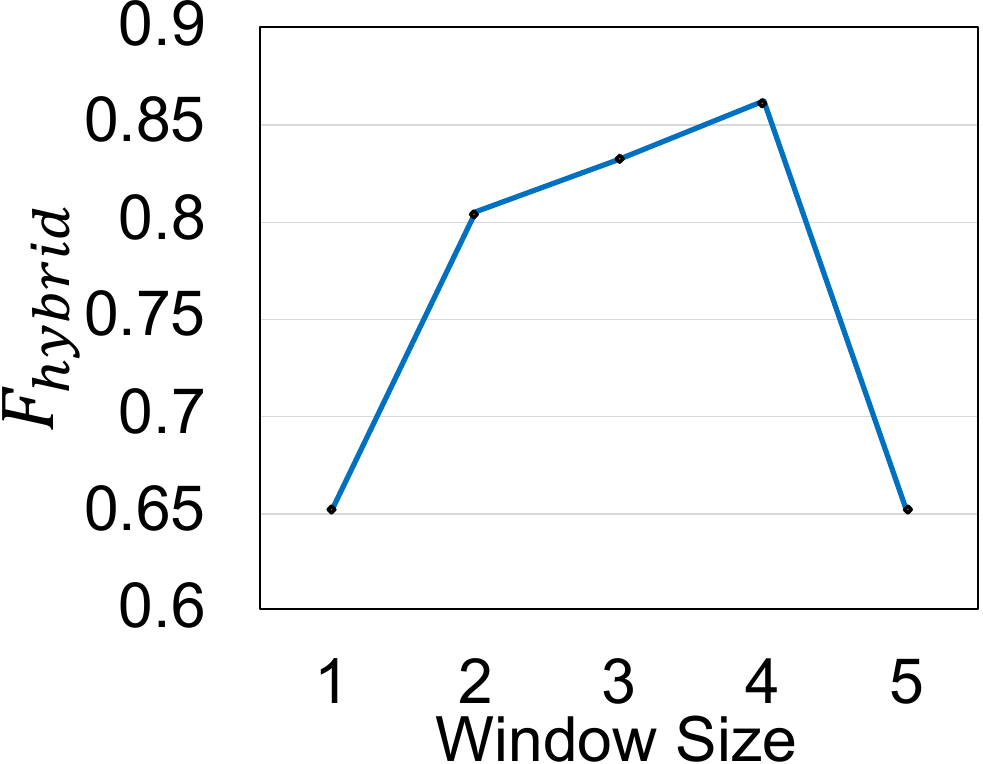} &
    \includegraphics[width=0.16\textwidth]{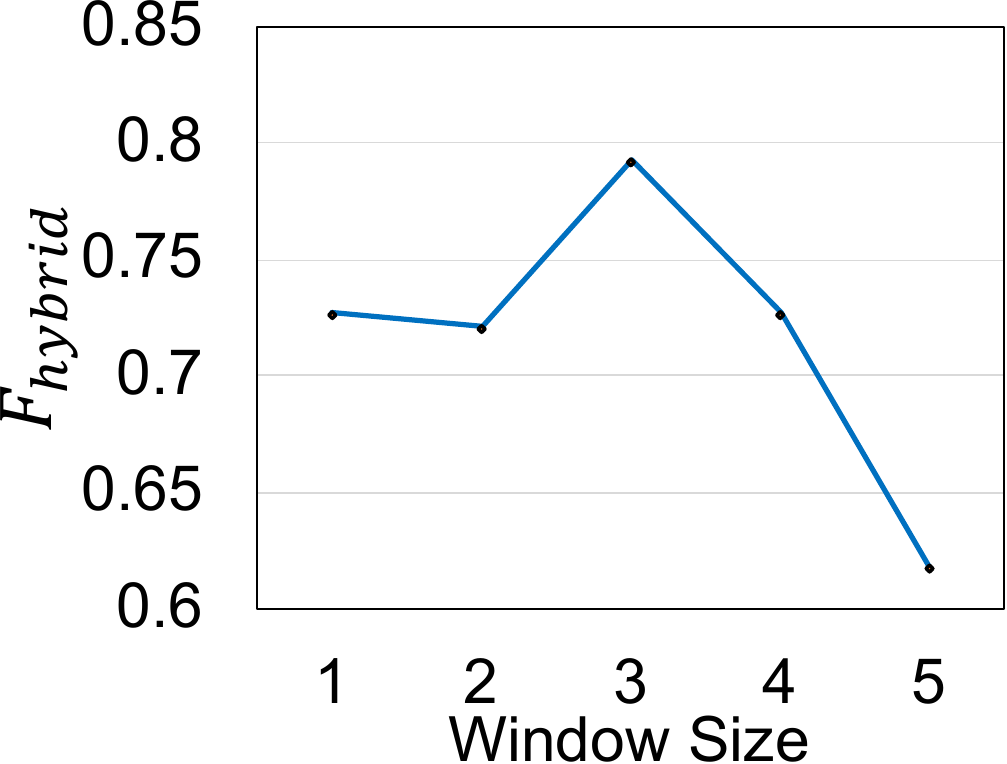} \\
    {\footnotesize (a) YouTube} & {\footnotesize (b) Clean Master} & {\footnotesize (c) Viber} & {\footnotesize (d) Ebay} & {\footnotesize (e) SwiftKey} & {\footnotesize (f) NOAA Radar} \\
    \multicolumn{6}{c}{{\small (2) Window Size.}} \\
    \includegraphics[width=0.16\textwidth]{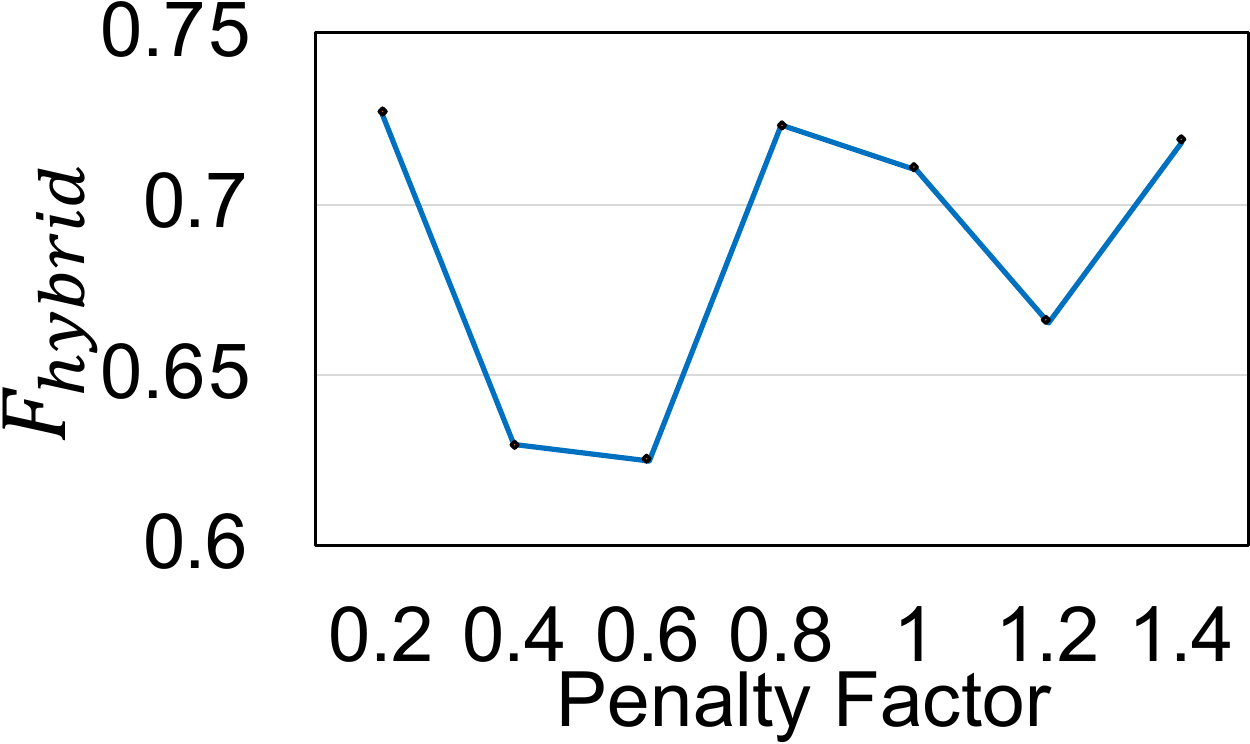} &
    \includegraphics[width=0.16\textwidth]{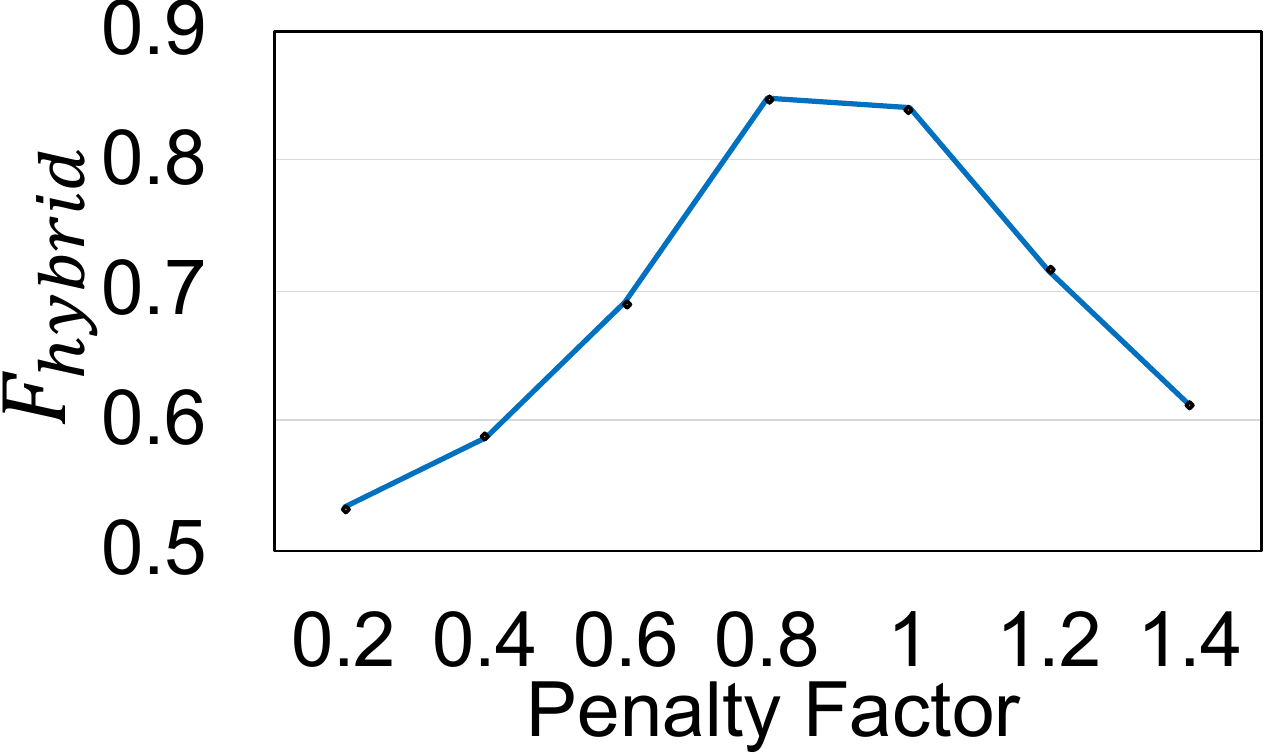} &
    \includegraphics[width=0.16\textwidth]{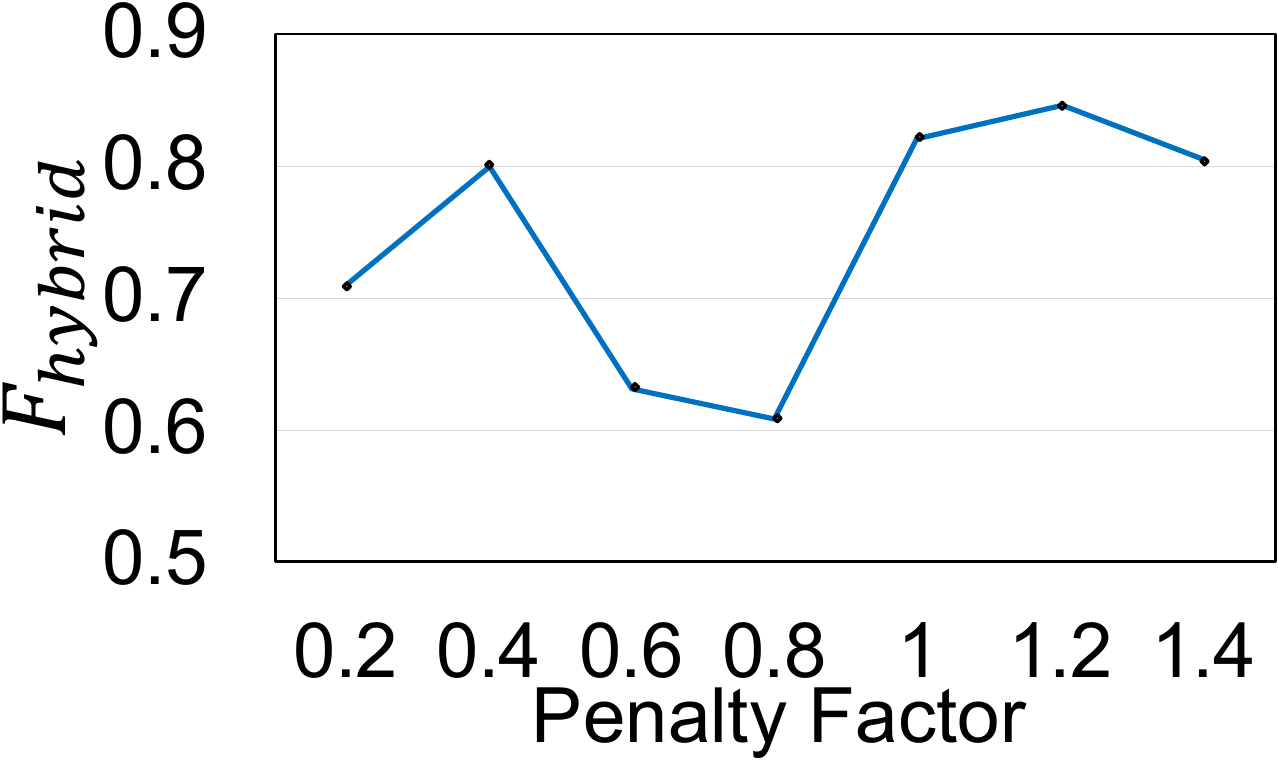} &
    \includegraphics[width=0.16\textwidth]{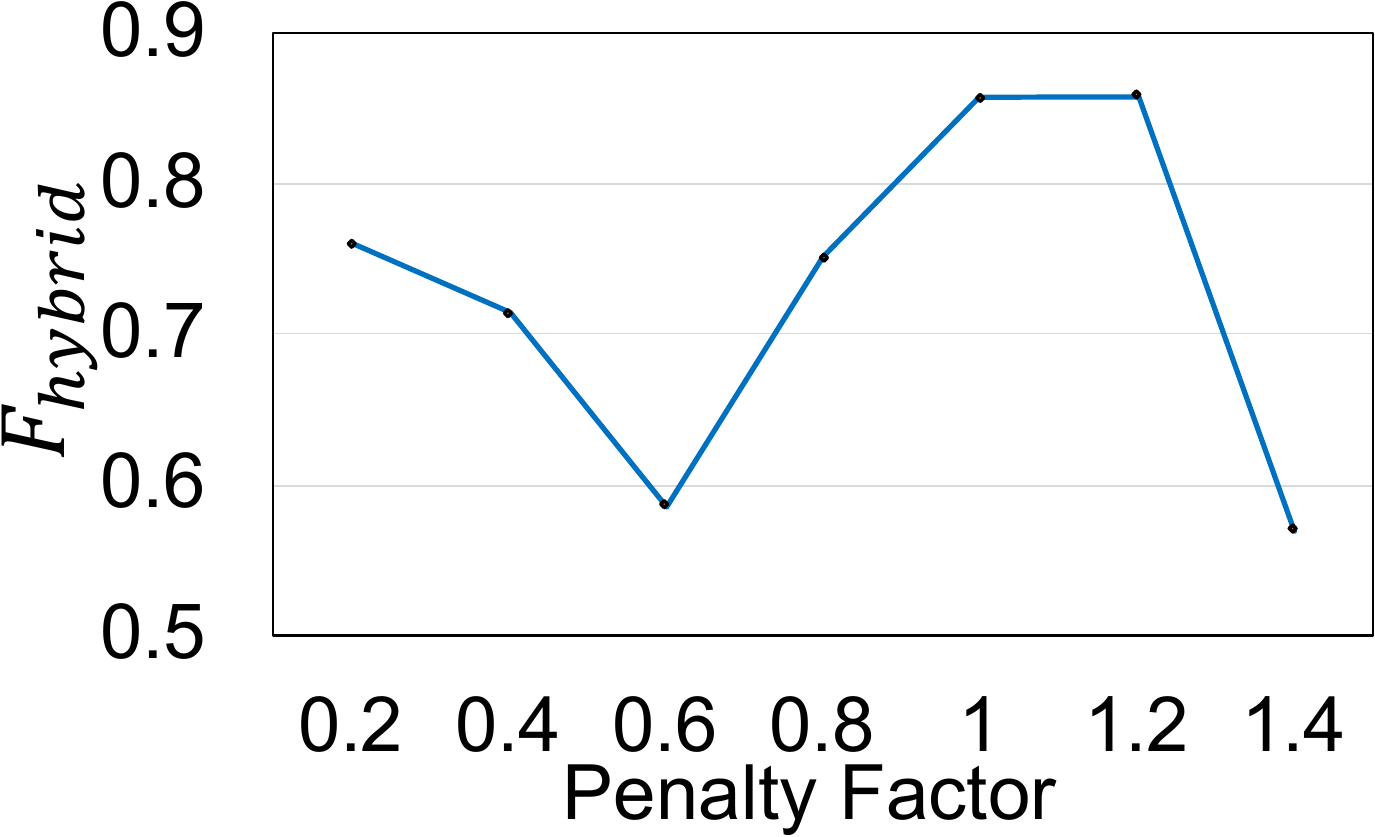} &
    \includegraphics[width=0.16\textwidth]{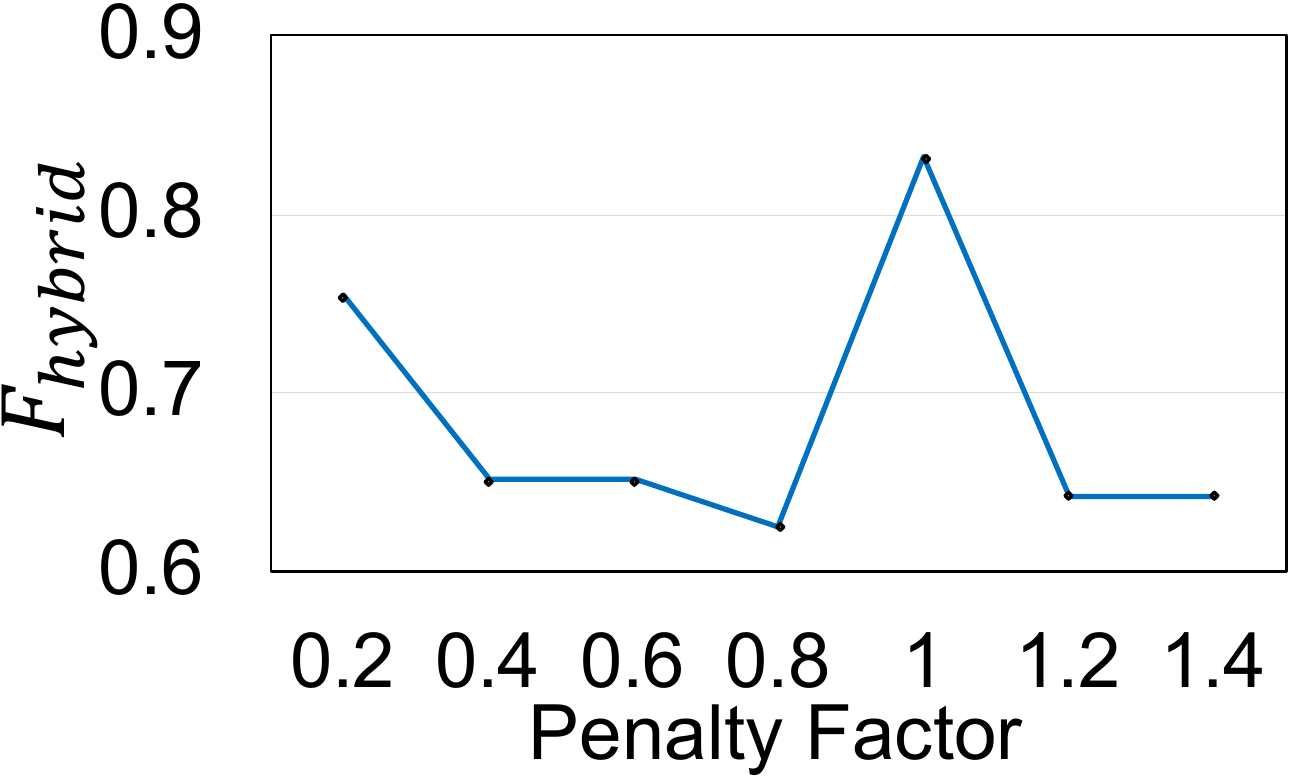} &
    \includegraphics[width=0.16\textwidth]{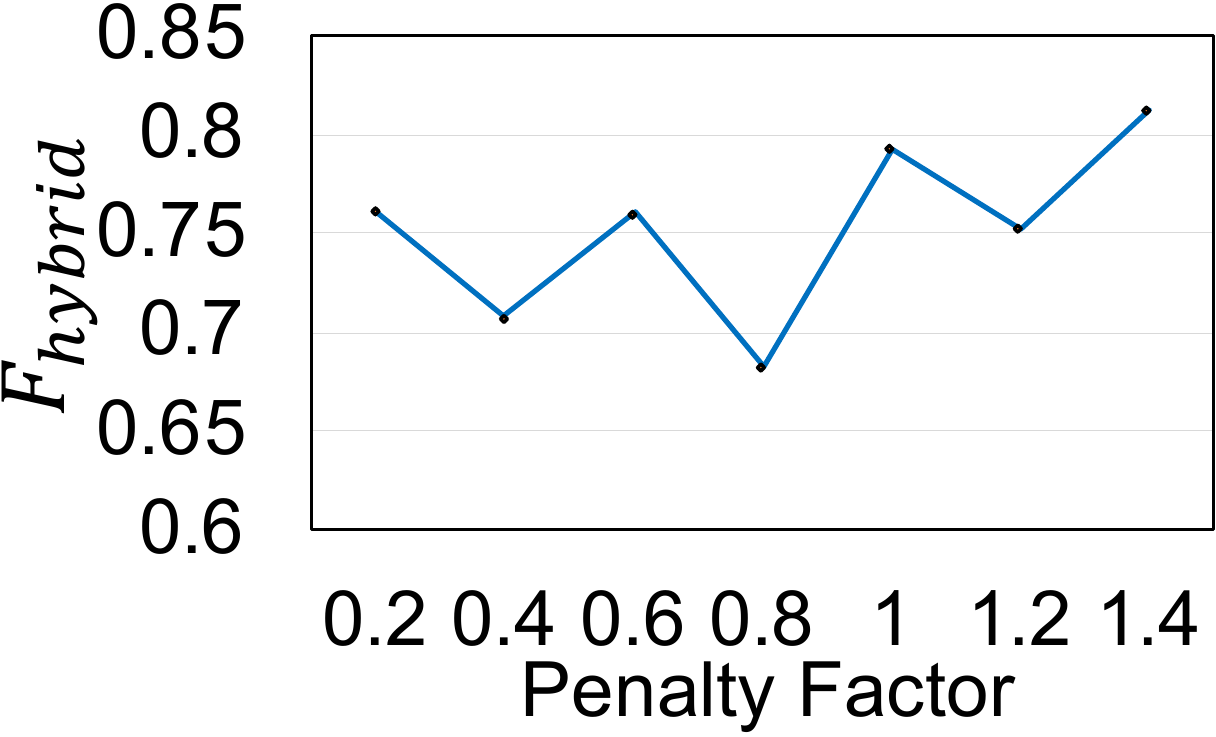} \\
    {\footnotesize (a) YouTube} & {\footnotesize (b) Clean Master} & {\footnotesize (c) Viber} & {\footnotesize (d) Ebay} & {\footnotesize (e) SwiftKey} & {\footnotesize (f) NOAA Radar} \\
    \multicolumn{6}{c}{{\small (3) Penalty Factor.}} \\
    \includegraphics[width=0.16\textwidth]{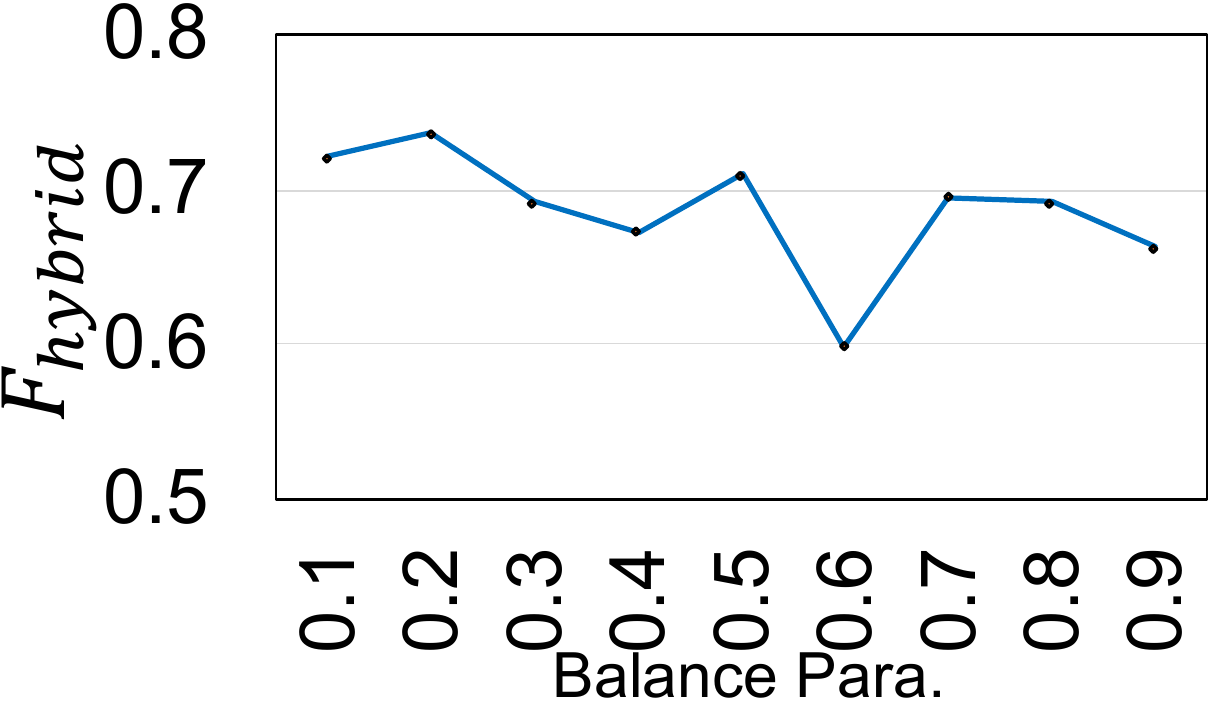} &
    \includegraphics[width=0.16\textwidth]{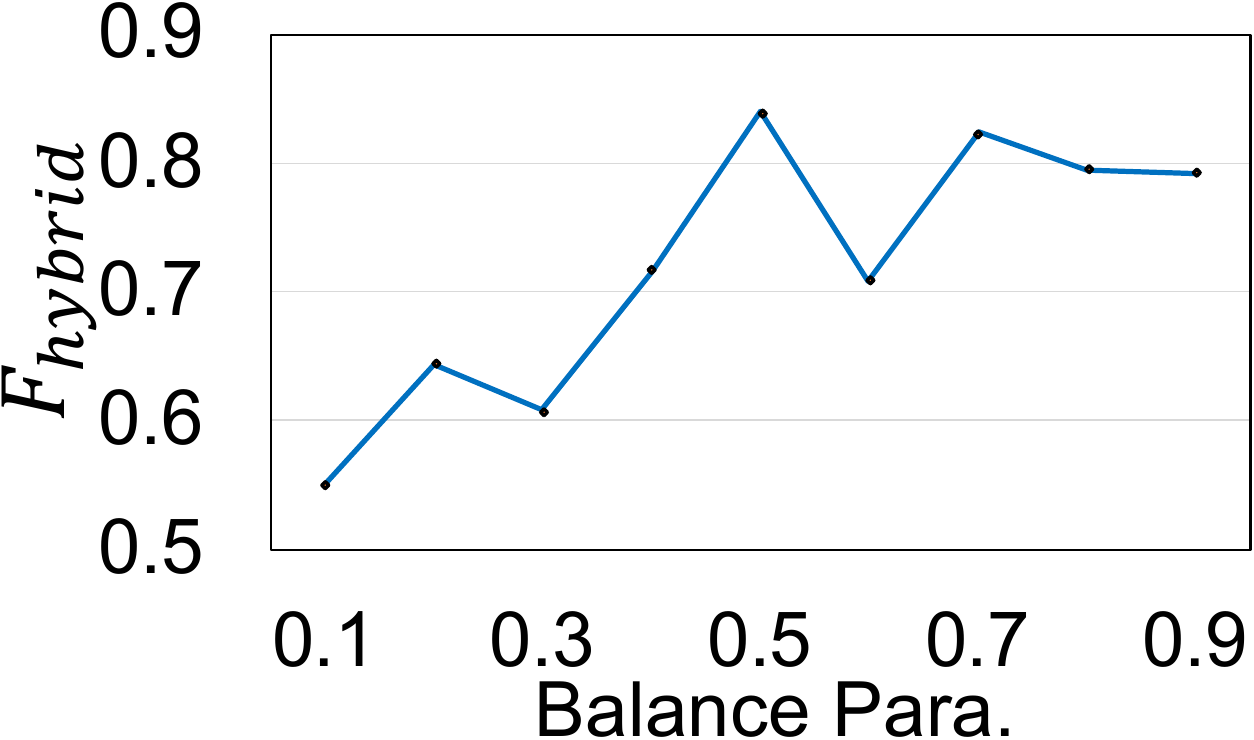} &
    \includegraphics[width=0.16\textwidth]{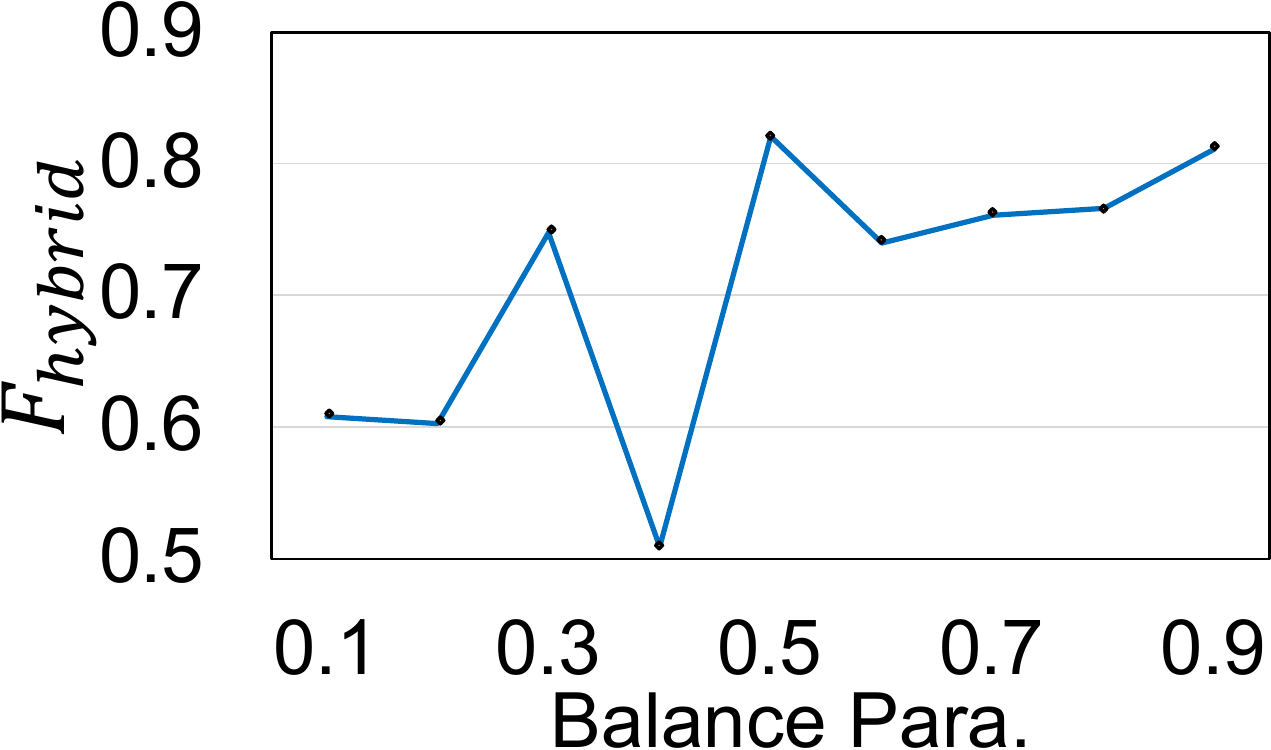} &
    \includegraphics[width=0.16\textwidth]{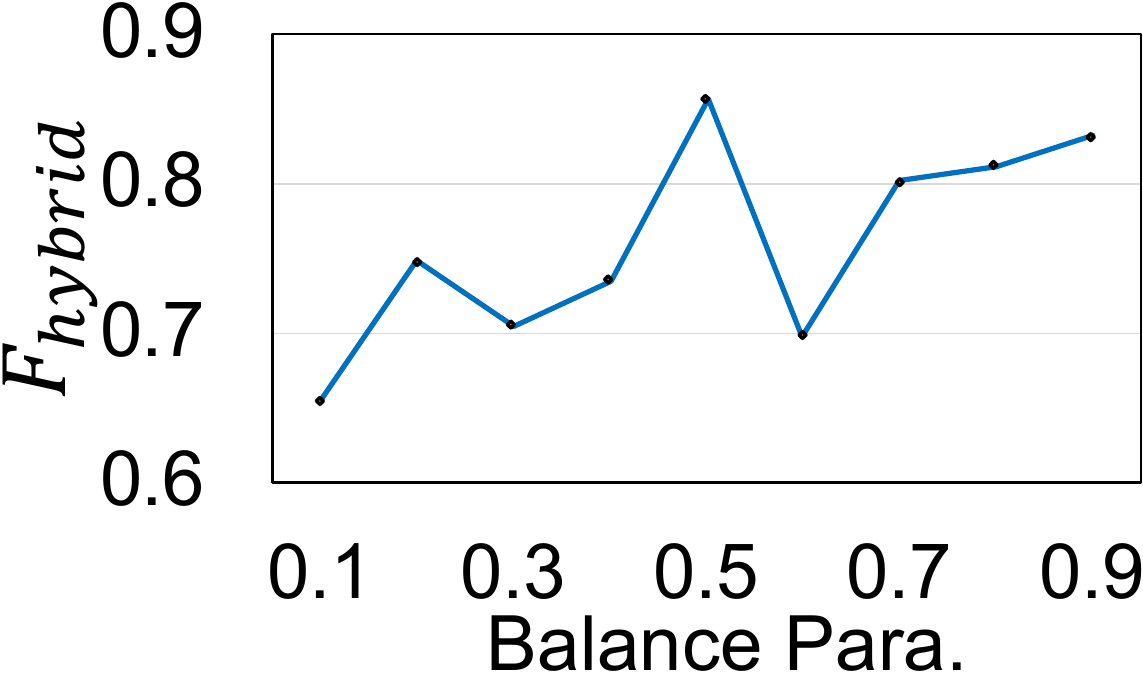} &
    \includegraphics[width=0.16\textwidth]{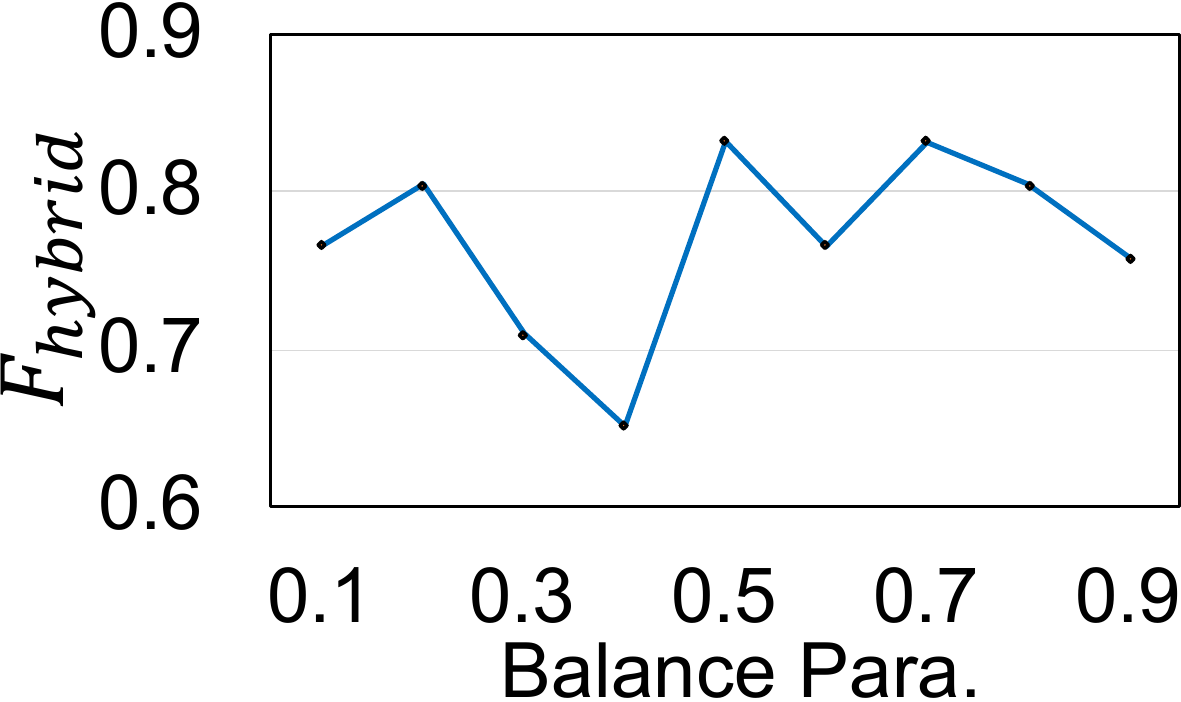} &
    \includegraphics[width=0.16\textwidth]{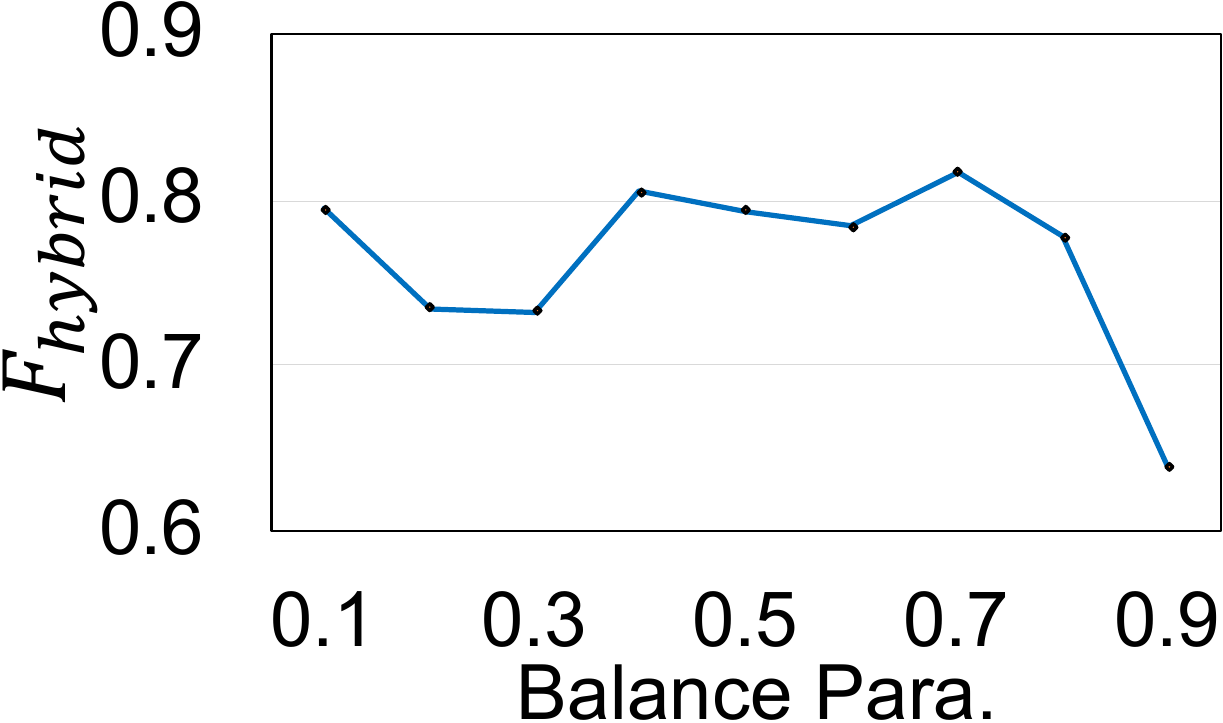} \\
    {\footnotesize (a) YouTube} & {\footnotesize (b) Clean Master} & {\footnotesize (c) Viber} & {\footnotesize (d) Ebay} & {\footnotesize (e) SwiftKey} & {\footnotesize (f) NOAA Radar} \\
    \multicolumn{6}{c}{{\small (4) Balance Parameter.}} \\
    \end{tabular}
    \caption{Impact of different parameters on the $F_{hybrid}$.}
	\label{fig:parameter}
\end{figure*}

As can be seen in Figure~\ref{fig:parameter} (1), the $F_{hybrid}$ score curves created by varying topic numbers are not consistent among the apps. For some apps such as YouTube, Clean Master, and Ebay apps, larger topic numbers can achieve better performance. However, for the SwiftKey and NOAA Radar apps, smaller topic numbers are preferred. This may be because the YouTube, Clean Master, and Ebay apps have relatively larger review volumes than the SwiftKey and NOAA Rader apps in the collected dataset, so more topics may exist. To better balance the precision and recall, we set the topic number as 13 during experiments.

\subsubsection{Window Size.}
According to Figure~\ref{fig:parameter} (2), the performance varies along with different window sizes. On the whole, the trends are analogous to an inverted ``U'' shape, such as the Viber, SwiftKey, and NOAA Radar apps. Such a phenomenon is reasonable since the topic distributions of the current version strongly rely on those of the previous versions within the window size. Smaller window sizes render the topic distributions of current versions more sensitive and unstable. Although larger window sizes can weaken the sensitiveness, they may also lack the sensitivity to emerging issues. We set window size $\omega$ as three since the setting can bring relatively better performance on the studied apps (indicated in Figure~\ref{fig:parameter}).

\subsubsection{Penalty Factor.}
As shown in Figure~\ref{fig:parameter} (3), an approximately inverted ``U'' shape can also be observed in most apps, such as the Clean Master, Ebay, and SwiftKey apps. Smaller penalty values may lead the ranked sentences to not distinguish the two topics well, and thereby prioritize similar sentence labels for the topics. In this way, the issues in sentences would not be able to cover all the emerging issues. However, larger penalty values may cause the label prioritization to put more weights on the distinguishability instead of the semantic similarity between the labels and topics, so the sentence labels may not well represent the meanings of current topics. We choose the penalty factor $\mu=1.0$ since the value presents almost the best performance on all the studied apps.

\subsubsection{Balance Parameter.}
The results under different balance parameters are illustrated in Figure~\ref{fig:parameter} (4). We can observe that generally higher balance parameters can lead to better performance for the studied apps, such as the Clean Master, Viber, and Ebay apps. However, such patterns are not applicable to the other apps. Since $m=0.5$ can achieve a good performance on our datasets, we set the balance parameter as 0.5 in our experiments.

\subsection{Manual Inspection}
In this paper, to automate the parameter tuning and result verification processes, we adopt a common semantic measurement metric~\cite{DBLP:conf/sigsoft/CambroneroLKS019}, \textit{i.e.}, cosine similarity. However, the automatic evaluation results might not exactly reflect practical performance. Manual evaluation is therefore needed to comprehensively evaluate the consistency between detected emerging issues and the ground truth. Since validating all the emerging issues would consume huge human effort, we choose the YouTube app, which contains the most versions among the studied apps (accounting for 37.1\% of all the app versions), for manual verification. The first two authors examined the detected emerging issues version by version with extensive discussions to reach a unanimous score. The results are shown in Table~\ref{tab:manual}. As can be seen, comparing manual inspection results with the scores computed by cosine similarity, the differences range from -5.2\% to +4.1\%. Also, in terms of the $F_{hybrid}$ scores, there is only a small disparity (around -1\%). Thus, using cosine similarity could be regarded as a reliable way to alleviate the labor burden and time consumed in both parameter tuning and verification. The effectiveness of MERIT is consequently confirmed.

\begin{table}[h]
	\center
	\caption{\yun{The results of manual evaluation on the YouTube dataset. The values inside brackets indicate the fluctuation ranges comparing to the scores computed by cosine similarity. Here, the comparison is based on the sentence-level issue representations.}}
	\label{tab:manual}
	\scalebox{0.9}{
	\begin{tabular}{c|c|c|c}
		\hline
		\hline
		Method & $Precision_{E}$ & $Recall_{L}$ & $F_{hybrid}$ \\\hline
		IDEA & 0.576 (-0.052) & 0.673 (+0.007) & 0.621 (-0.015) \\
		MERIT & 0.618 (-0.049) & 0.801 (+0.041) & 0.698 (-0.012)\\
		\hline
		\hline

	\end{tabular}
}
\end{table}

\subsection{Survey on Industry Practitioners}
To further demonstrate the effectiveness and practicability of our work, we conduct a user study among 44 industry practitioners in large IT companies including Alibaba and Tencent, with 19 developers (43.2\%), nine data analysts (20.5\%), six test engineer (13.6\%), two product managers (4.5\%), and 10 from other positions (22.7\%). Around 80\% of the participants have more than one year software engineering experience. The developer survey is conducted through an online questionnaire, which consists of five questions: two questions on participants' background and three questions for understanding their attitude towards the practicability of MERIT. During the survey, we validate the practicability of MERIT in terms of three aspects: acceptability of the provided emerging issues, preference of higher accuracy but with more time consumption, and willingness of applying such a tool into their development pipeline. Each aspect is rated on a 1-4 Likert scale (4 for agreement, 3 for mild agreement, 2 for mild disagreement, and 1 for disagreement).

\subsubsection{Acceptability of the provided emerging issues}
We survey the participants about their opinions on the presented descriptions of the emerging issues, by providing one official changelog example of YouTube and corresponding detected emerging issues, \textit{i.e.}, topic labels in phrase and sentence. The survey results indicate that all the interviewees (100\%) agree that the provided issues are acceptable, among which 15 (34.1\%) of them are fully in favor of the usefulness of the issue descriptions.

\subsubsection{Preference of higher accuracy but with more time cost}
To investigate on developer's opinions regarding the performance of MERIT, especially whether the higher accuracy is worth the more time cost, we present the performance of two model examples and ask participants to choose the preferred one. The two models are: Model A can process 200 reviews in one second and obtains 60\% accuracy; while model B achieves 80\% accuracy but can only process 125 reviews per second. According to the survey, 27 (61.4\%) of the interviewees prefer model B which achieves a higher accuracy but with a lower processing speed, and eight interviewees (18.2\%) consider both models to be acceptable. Only nine survey respondents (20.5\%) chose model A. The results indicate that industry practitioners possibly prefer MERIT than the baselines regarding the performance.

\subsubsection{Willingness of applying MERIT into industry}
We collect participants' opinions of whether they are willing to employ or recommend developers to employ our tool MERIT. According to the survey, around 95.5\% of the interviewees express that they would possibly use MERIT in their development pipelines, and 34.1\% fully approve of its adoption in practice and think MERIT can reduce the effort in manual analysis. The results demonstrate the potential benefit of MERIT to developers.

\vspace{5mm}
\subsection{Threat to Validity}
First, our model evaluation is based on the six subject apps in~\cite{gao2018online}, which may not guarantee the generalization of the findings. We pick the dataset used in~\cite{gao2018online} to allow for fair comparison. Second, app versions with few user reviews can impact the performance of MERIT. Since small datasets can be easily analyzed manually, MERIT is targeted for automatic analysis of large review datasets. Also, MERIT"s good performance on different quantities of user reviews (on average 523$\sim$6,332 reviews per version) show that MERIT would well adapt to different review sizes. \yun{Third, the 500 opinion words manually labeled during polarity word preparation procedure may not be the optimal opinion words for inferring the sentiment associated with each topic. To mitigate the threat, we randomly selected the words weighted by their frequencies, so the words with higher frequencies are more likely to be selected and labeled. Also, we ensure the sample corresponds to a statistically significant proportion of the whole opinion lexicons. In practice, app developers can choose a different opinion word set for emerging issue detection. How to select an optimal set of opinion words for better sentiment inference can be future work. Fourth, the changelogs and cosine similarity measurement that we adopt for evaluation may not accurately reflect the practical performance. We mitigate this threat by manually validating the results on a sample of reviews and demonstrating that the results reported using cosine similarity are consistent with those obtained via manual inspection.}

\yun{Moreover, in this study, we only combine the sentiment characteristic of app reviews into MERIT while other factors such as device types that may be helpful for emerging issue detection are not involved. Future studies should broaden the set of features used to characterize app reviews in our study and investigate the impact of different characteristics on the performance of emerging issue detection. In addition, there may be alternative approaches to combine topics and sentiment for emerging issue detection, e.g., implementing a two-step pipeline where extracting negative review sentences or paragraphs is the first step, and modeling topics of the negative texts is the second step. We leave implementation and evaluation of these alternatives to future work.} Finally, MERIT shares the same limitation with the adopted topic modeling approach~\cite{DBLP:conf/www/YanGLC13}, \textit{i.e.}, the number of topics should be determined initially. This limitation is brought by the unsupervised nature of the approach. There are studies~\cite{DBLP:conf/pakdd/ArunSMM10,zhao2015heuristic,DBLP:journals/corr/abs-1708-01677} on automatically identifying the optimal topic number, but they are not easy to be adapted to online topic modeling approaches, which is the core of our proposed framework. How to efficiently discover the optimum topic numbers for online topic models can be regarded as a challenging and interesting work for future research.

%  Because the optimal topic numbers of different review corpora are different, and we could not simply choose the optimum value on one corpus as the .

\section{Related Work}\label{sec:literature}
We discuss two threads of studies that inspire our work: App review analysis and emerging topic detection.

\subsection{App Review Analysis}
Since app reviews serve as an essential channel between users and developers, and provide rich information about app usage, the number of studies on user review analysis is on the rise~\cite{DBLP:journals/tse/MartinSJZH17}. \yun{Recent research has leveraged Natural Language Processing and Machine Learning techniques to extract useful information from online app products to help developers realize, test, optimize, maintain, and release apps (see \textit{e.g.}, \cite{Palomba2017ICSE,Grano2018,DBLP:conf/kbse/UddinK17a,DBLP:conf/icse/0008ZBPL19}).} The major goal of these studies is to alleviate the burden of summarizing useful knowledge from a relatively huge quantity of unstructured texts. Here, we focus on the research that exploiting app reviews to facilitate the process of app maintenance and release.

A number of studies~\cite{platzer2011opportunities,DBLP:conf/re/MaalejN15,DBLP:conf/ease/LicorishSK17,DBLP:journals/software/KhalidSNH15} categorize user reviews based on their sentiment (e.g., either praise or complaint) and general topics (e.g., bug report or feature request). Di Sorbo et al.~\cite{di2016would} presented an approach called SURF to further classify reviews into fine-grained topics (e.g., GUI and security). Based on the categorized reviews, Gu and Kim~\cite{DBLP:conf/kbse/GuK15} applied aspect opinion mining and sentiment analysis to find the most popular features of an app. Although they can present the rating changes of one app feature over time, the ratings are tracked based on feature words instead of topics. Moreover, their work does not establish the relation of features with star-ratings~\cite{noei2019too}. \yun{Besides, Islam and Zibran~\cite{DBLP:conf/msr/IslamZ17} and Calefato et al.~\cite{DBLP:journals/ese/CalefatoLMN18} design sentiment analysis tools specific to software development.}

Topic modeling is widely used in different domains, and interesting results have been inferred~\cite{DBLP:conf/nips/HoffmanBB10,DBLP:conf/sose/GaoXHZ15}. Consequently, some researchers rely on topic modeling technique~\cite{DBLP:conf/nips/BleiNJ01,DBLP:conf/www/YanGLC13} to analyze user reviews. Iacob and Harrison~\cite{DBLP:conf/msr/IacobH13} and Guzman and Maalej~\cite{guzman2014users} applied LDA to extract app features. Chen et al.~\cite{chen2014ar} adopted LDA to capture the topic distribution of each user review, based on which they prioritized useful user reviews to developers. Fu et al.~\cite{DBLP:conf/kdd/FuLLFHS13} analyzed the changes in the review number associated with each topic over time. Noei et al.~\cite{noei2019too} used LDA to determine the key topics of user reviews for different app categories. Gao et al.~\cite{DBLP:conf/sose/GaoXHZ15,DBLP:conf/issre/GaoWHZZL15} resorted to topic modeling methods for prioritizing app issues. None of the papers mentioned above have considered the sentiment changes of topics along with time or exploited the changes to detect emerging app issues.

Our previous work~\cite{gao2018online} is the most recent study focusing on tracking app issues along with release versions. Specifically, the IDEA proposed in~\cite{gao2018online} analyzed issue changes along with app versions using online topic modeling during which the emerging app issues are identified. Another of our recent work~\cite{gao2019diver} also aimed at detecting emerging app issues but mainly during the beta testing periods. Although the IDEA model performs well on the studied apps, the proposed model still meets several limitations as discussed in Section~\ref{sec:introduction}. 

Nayebi \textit{et al.} investigate app updating frequency and its impact~\cite{DBLP:conf/wcre/NayebiAR16,nayebi2016analysis}. They find that users prefer to install apps that were updated more recently and less frequently. Thus, frequent updates are not always considered positively in practice. Updating frequencies should also be carefully determined. Determining the sweet spot for app update frequency is an important and an interesting research topic. As this is beyond the scope of the current paper, we will leave this for future work.

% ; while the DIVER in~\cite{gao2019diver} employs frequent pattern mining and clustering techniques with both time and version dimensions considered. Although the DIVER is demonstrated to be effective in the industry scenario, its issue tracking is word-based, where only the words with abnormal increase are considered, and might miss the abnormal topics composed of the words with normal increase. Besides, both studies do not explicitly involve user sentiment, which can indicate whether a topic is complained or praised by users.

% Since word embedding (i.e., word2vec)~\cite{DBLP:journals/corr/abs-1301-3781} has proven to be effectively capture the semantic and syntactic relations between words, it is usually adopted to retrieve the 

\subsection{Emerging Topic Detection}
An event\footnote{An \textit{event} in social media corresponds to an \textit{app issue} in the context of app reviews.}, in the context of social media, is an occurrence of interest in the real world which initiates a discussion on the event-associated \textit{topic} on social media platforms, either soon after the occurrence or, sometimes, in anticipation of it. The emerging event detection approaches can be based on term interestingness~\cite{DBLP:conf/chi/MarcusBBKMM11}, incremental clustering~\cite{DBLP:journals/jocs/KaleelA15}, or topic modeling~\cite{DBLP:journals/concurrency/LiWTZY16}, etc. For example, Li et al.~\cite{DBLP:conf/cikm/LiSD12} identified emerging events from Twitter by first selecting top bursty words and then conduct word clustering, which is a term-interestingness-based approach. A comprehensive survey on emerging topic detection approaches can be found in Hasan \textit{et al.}'s work~\cite{DBLP:journals/jis/HasanOS18}. Different from texts on social media, each review is specific to one app version, and generally shorter in length~\cite{amazonreviews,DBLP:journals/jss/Genc-NayebiA17}, which renders app review mining a more challenging task, specialized to software engineering context. However, the existing studies~\cite{DBLP:journals/jis/HasanOS18} in machine learning field either do not involve automatic labeling of topics or do not consider the short length nature of input texts, so directly applying them into our app review scenario will not be optimal.

Thus far in the app review analysis literature, term interestingness~\cite{gao2019diver,DBLP:conf/kbse/VuPNN16,DBLP:conf/kbse/VuNPN15} and topic modeling methods~\cite{gao2018online} have been widely used.

\textit{The term-interestingness-based methods} rely on tracking the terms likely to be related to being an event, and are usually followed with clustering methods. Various approaches are proposed to determine the interestingness score\footnote{The \textit{interestingness score} refers to the possibility of a term to be related to an emerging event.} of each term. For example, Minh Vu et al.~\cite{DBLP:conf/kbse/VuPNN16,DBLP:conf/kbse/VuNPN15} first grouped the keywords using clustering algorithms and then determine the emergent clusters based on the occurrence frequencies of the keywords in each cluster. 

% However, the term-interestingness-based methods heavily rely on word frequencies and may miss the emerging issues which are described using various words and 

% Gao et al.~\cite{gao2019diver} identify emergent terms leveraging the usage of the terms in previous time intervals and release versions. A clustering technique is then employed to group the terms based on their semantic similarity. The concrete meanings of these emerging groups are finally interpreted manually by developers. Different from Gao et al.~\cite{gao2019diver}, 

\textit{The topic-modeling-based methods} associate each document with a probability distribution over various latent topics and track the topic distributions over time. For example, our previous work~\cite{gao2018online} proposed an online topic modeling approach to infer the topic distributions of user reviews along with app releases. The emerging topics are identified based on their differences with the corresponding topics in previous time slices.

% to infer the hidden semantic structures of a collection of documents

The term-interestingness-based methods can be regarded as a down-top model (\textit{i.e.}, from word to topic), while the topic-modeling-based methods are top-down (\textit{i.e.}, from topic to word). Since the abruptness of one topic does not indicate that all the words belonging to the topic show bursty trends, term-interestingness-based models may generate true negatives due to missing bursty words. Thus, our proposed model is based on the topic modeling approach.

% One advantage of the topic-modeling-based methods over the term-interestingness-based methods is that 

% the former one can represent semantically-similar words with probability vectors that are closely located to each other in a dense-dimension space.  

%\IEEEpubidadjcol

\section{Conclusion and Future Work}\label{sec:conclusion}
To ensure good user experience and maintain high-quality apps, identifying emerging issues in a timely and accurate manner is critical. In this paper, we propose a novel topic-modeling-based framework named MERIT for detecting emerging issues by analyzing online app reviews. MERIT improves the state-of-the-art method by better modeling of short review texts, jointly modeling topics and sentiment, and using word embeddings to better interpret topics. Extensive experiments verify the effectiveness and efficiency of our proposed framework, MERIT. In the future, we will conduct evaluations using a larger dataset and deploy the model with our industry partners.

% \appendices

% % you can choose not to have a title for an appendix
% % if you want by leaving the argument blank
% \section{}

% % use section* for acknowledgment
% \ifCLASSOPTIONcompsoc
%   % The Computer Society usually uses the plural form
%   \section*{Acknowledgments}
% \else
%   % regular IEEE prefers the singular form
% \fi


\ifCLASSOPTIONcaptionsoff
  \newpage
\fi

\bibliographystyle{IEEEtran}
\bibliography{sigproc}   % name your BibTeX data base

% \end{IEEEbiography}


% % insert where needed to balance the two columns on the last page with
% %\newpage

\end{document}